\documentclass[12pt,preprint]{aastex}

\usepackage{lscape}

\usepackage[usenames,dvips]{color}
\usepackage{graphicx}
\shorttitle{X/$\gamma$-ray emissions from pulsar wind of PSR B1259-63}
\shortauthors{Takata J. \& Taam E.R.}

\begin{document}


\title{Probing the Pulsar Wind in the $\gamma$-ray Binary System PSR B1259-63/SS 2883}


\author{
Jumpei Takata\altaffilmark{1,2}
\email{takata@tiara.sinica.edu.tw}
\and
Ronald E. Taam\altaffilmark{1,2,3}
\email{r-taam@northwestern.edu}
}
\altaffiltext{1}{Academia Sinica Institute of Astrophysics and Astronomy-TIARA, P.O. Box 23-141, Taipei, 10617 Taiwan}
\altaffiltext{2}{Academia Sinica Institute of Astrophysics and Astronomy/National Tsing Hua University-TIARA, Hsinchu Taiwan}
\altaffiltext{3}{Northwestern University, Department of Physics and Astronomy, 2131 Tech Drive,
Evanston, IL 60208}


\begin{abstract}
The spectral energy distribution from the X-ray to the very high energy regime 
($>100$ GeV) has been investigated for the $\gamma$-ray binary system PSR 
B1259-63/SS2883 as a function of orbital phase within the framework of a simple 
model of a pulsar wind nebula. The emission model is based on the synchrotron 
radiation process for the X-ray regime and the inverse Compton scattering process
boosting stellar photons from the Be star companion to the very high energy 
(100GeV-TeV) regime.  With this model, the observed temporal behavior can, 
in principle, be used to probe the pulsar wind properties at the shock as a 
function of the orbital phase. Due to theoretical uncertainties
 in the detailed 
microphysics of the acceleration process and the conversion of magnetic energy 
into particle kinetic energy, the observed X-ray data for the entire orbit are 
fit using two different methods. In the first method, the magnetization 
parameter and the 
Lorentz factor of the wind at the shock are allowed to vary for a given power law index 
characterizing the accelerated particles at the shock. In this case, the observed 
photon index $\sim1.2$ in the 1-10~keV energy band near periastron passage can 
be understood provided that (a) the electron energy distribution is described by 
a broken power law and (b) there is a break at an energy of about $8\times 10^{6}$ 
in units of the electron rest mass energy. In the second method, the magnetization 
parameter and the power law index are varied for a fixed Lorentz factor.  Here, the 
photon index $\sim 1.2$ can result from a particle distribution described by a 
power law index of $\sim 1.5$.  The calculated emission in the energy band 
corresponding to 10~MeV-1~GeV from the shocked pulsar wind indicate that these two 
cases can be distinguished by future $Fermi$ observations near the periastron.
It is also found that the emission
from the unshocked wind could be detectable by the Fermi telescope near periastron 
passage if most of the kinetic energy of the flow is carried by particles with 
Lorentz factors of $\Gamma\sim 10^5$.
\end{abstract}





\section{Introduction}

The study of cosmic sources emitting electromagnetic radiation at very high 
energies in the TeV range has been greatly facilitated by the capability of ground 
based Cherenkov telescopes.  With access to this new observational window, a class 
of binary systems has been recently revealed and characterized by the emission of high 
energy radiation. Specifically, very high energy $\gamma$-rays were detected from 
three binary systems in the Galaxy; namely, PSR B1259-63 (Aharonian et al. 2005), 
LS I $\mathrm{+61^{o}}$~303 (Albert al et. 2006), and LS 5039 (Aharonian et al. 
2006). These $\gamma$-ray binaries are high mass X-ray binary systems 
composed of a compact object (a neutron star or black hole) in orbit around a high 
mass main sequence-like star (Dubus 2006b). Since very high energy emission implies 
the existence of charged particles in the ultra-relativistic regime, these $\gamma$-ray 
binaries can be viewed as cosmic laboratories for studies of particle acceleration.

The orbital properties of the $\gamma$-ray binary system associated with the Be star 
SS 2883 are unusual in that it has a large eccentricity ($e=0.87$) and a wide orbit 
with period of 1287 days.  Particularly noteworthy is the presence of a young radio 
pulsar companion (PSR B1259-63) characterized by a spin period of $P=47.76$ ms and 
a spin down energy of $\dot{E}_{sp}=8\times 10^{35}$ ~erg s$^{-1}$.  This system had 
been known as a source of non-pulsed and non-thermal emission in both the radio and 
X-ray energy bands (Johnston et al.  2005; Uchiyama et al. 2009), and the recent 
observations based on H.E.S.S.  revealed the existence of very high-energy $\gamma$-rays 
(above 380~GeV) at orbital phases close to periastron passage (Aharonian et al. 2005). 
 
PSR B1259-63/SS 2883 has also been observed to exhibit large temporal variations in its 
emission and spectrum interpreted as resulting from variations in the orbital separation 
of the components in the system. The observational sampling of the orbit, however, is 
not uniform among the different wave-bands.  Although the X-ray emission is measured 
throughout the orbit (Hirayama et al. 1996; Chernyakova et al. 2006; 
Uchiyama et al. 2009), the radio and 
the TeV emissions have been detected only during orbital phases near periastron passage 
(Johnston et al 2005; Aharonian et al 2005). The photon index characterizing the 
spectral energy distribution in the 2-10~keV band is found to vary with the orbital 
phase, ranging from about 1.2 to 2. 
 
The origin of this high energy emission is likely related to the interaction of the 
pulsar wind of PSR B1259-63 with the outflow from the Be star (Tavani \& Arons 1997). Their 
interaction results in the formation of a termination shock where the dynamical 
pressures of the pulsar wind and the stellar wind of the Be star are in balance. 
The shocked pulsar wind particles can emit non-thermal photons over a wide range 
of energies via the synchrotron radiation and inverse Compton processes. This 
picture for the origin of the non-thermal emission is very similar to that developed 
for the emission from isolated pulsars, such as the Crab pulsar (Kennel \& Coroniti 1984a,
b; Atoyan \& Aharonian 1996). However, in this interpretation, the high energy 
emission regions in $\gamma$-ray binary systems lie close to the pulsar occurring 
within the orbital separation of the two components with the emission taking place 
in regions characterized by greater magnetic fields than those of their isolated 
counterparts. Specifically, the distance to the termination shock from the pulsar, $r_s$, 
in the PSR B1259-63/SS2883 system is in the range $r_s\sim 0.1-$1~AU, and 
the strength of the magnetic field is in the range $B\sim 10^{-3}\sim 0.1$~G, while 
$r_s\sim 0.1$~pc and $B\sim 100~\mu$G for isolated pulsars.  Therefore, $\gamma$-ray 
binaries provide a unique laboratory to probe the physics of the pulsar wind in a 
different regime distinct from those studies in isolated pulsars.

Within the framework of the interacting winds model, the observed temporal 
behavior of the flux and the photon index is caused by variations in the 
physical conditions at the termination shock of the pulsar wind. Because the 
distance of the shock from the pulsar is a function of the orbital phase, the 
physical properties of the pulsar wind can be probed as a function of the 
radial distance from the pulsar. The properties of the pulsar wind after 
the shock are determined by the wind magnetization parameter, $\sigma$, and the 
spin down luminosity $\dot{E}_{sp}$ of the pulsar.  The $\sigma$ parameter  is  
defined by the ratio of the magnetic energy to kinetic energy  upstream 
of the shock as

\begin{equation}
\sigma=\frac{B_1^2}{4\pi \Gamma_1 u_1 n_1m_ec^2}.
\label{sigma}
\end{equation}
Here $B_1$ and $n_1$ are the pre shock magnetic field in the shock frame 
and proper number density of the electrons 
and positrons respectively.  In addition, $u_1$ and $\Gamma_1=\sqrt{1+u_1^2}$ 
are the dimensionless radial four velocity and   the Lorentz factor of the 
unshocked flow, respectively, and $m_e$ is the rest mass of the electron and 
$c$ is the speed of light. The proper number density $n_1$ and the magnetic 
field $B_1$ upstream of the shock are taken from Kennel \& Coroniti 
(1984a, b) as 
\begin{equation}
n_1=\frac{\dot{E}_{sp}}{4\pi u_1\Gamma_1 r_s^2m_ec^3(1+\sigma)}
\label{number}
\end{equation}
and 
\begin{equation}
B_1^2=\frac{\dot{E}_{sp}\sigma}{r_s^2c(1+\sigma)}, 
\label{b1}
\end{equation}
respectively, where $r_s$ is the distance of the shock from the pulsar.

At the shock, the kinetic energy of the pulsar wind is converted into 
the internal energy  of the wind, and the distribution of downstream 
 particles  is assumed to be described 
by a power law over several decades in energy.
 For isolated pulsars, small values of the $\sigma$ parameter
 have been inferred ($\sigma\sim 
0.003$) by Kennel \& Coroniti (1984a, b), suggesting 
a kinetic energy dominant flow, 
to explain, for example, the expansion velocity of the Crab pulsar wind.
It is known, however, that it is difficult to theoretically reproduce 
the kinetically dominated relativistic MHD flow (Bogovalov 1999), 
although some possible explanations for the small values for $\sigma$ have
 been discussed by 
Kirk \& Skj\ae raasen (2003) and Arons (2008).
The resolution of the $\sigma$ problem has yet to be clarified, as well as the 
acceleration mechanism at the shock. 
As an alternative approach, the study of the $\gamma$-ray binaries provides us 
with an astrophysical site to phenomenologically probe the $\sigma$ parameter 
of the pulsar wind at varying distances from the pulsar.  

The observed non-thermal emission from the radio to the very high energy 
band has been interpreted within the framework of a  leptonic model 
(Tavani \& Arons 1997; Kirk et al. 1999; Dubus 2006b) or in terms of a 
hadronic model (Neronov \& Chernyakova 2007). In the leptonic model,  
the electrons and positrons are accelerated at the shock front and 
distributed over several decades in energy by a single power law index.
The synchrotron emission from these particles produces non-thermal photons 
from the radio regime through the GeV $\gamma$-ray bands. On the other 
hand, the very high-energy ($>$ 100~GeV) photons are produced by 
the inverse Compton process whereby the electrons and positrons 
scatter  optical  photons from the Be star. In the hadronic model,
 the broad band non-thermal emission is produced as a result of collisions 
of protons accelerated at the shock front with the circumstellar disk.  The very 
high energy (TeV) emission is produced via the decay of pions, 
while the radio to GeV emission is attributed to the synchrotron and inverse Compton 
processes involving the low-energy electrons and positrons produced by the 
decay of the charged pions.

In this paper we re-investigate the emission characteristics of the PSR 
B1259-63/Be star SS2883 binary system in view of recent X-ray observational
data to place constraints on the properties of the pulsar wind by further 
developing the synchrotron and inverse-Compton emission model for the
 non-thermal emission. 
The observations reveal that the temporal behavior of the 
X-ray emission predicted 
by Tavani \& Arons (1997) is not consistent with the recent observational data.  
For example, the photon index $\alpha$ was predicted to vary between $\alpha\sim 
1.5-2$ in the orbit, while the observed photon index is in the range $\alpha\sim 
1.2-1.8.$ In addition, at the very high energy bands, there is a tendency for the 
flux to attain a minimum at the epoch close to the periastron as observed by H.E.S.S 
(Aharonian et al. 2005). Although TeV emission with the leptonic model has been studied 
(e.g. Kirk et al. 2005; Khangulyan et al. 2007; Sierpowska-Bartosik \& Bednarek 2008), 
the observed temporal behavior (in particular close to periastron) is not fully understood. From the theoretical perspective, we improve upon the study of Tavani \& Arons 
(1997) by including the full corrections to the Klein Nishina cross section, 
following Kirk et al. (1999) and Khangulyan et al. (2007),  since 
the inverse Compton process between the accelerated particles and the upscattered 
photons from the Be star occurs in a regime where such corrections can be important. 
Tavani \& Arons (1996) carried out seminal work on the X-ray emission, but
it is highly desirable to describe the temporal behavior of both the 
X-ray and very high-energy emission together.

The observed photon index below 1.5 in the 1-10~keV energy band is one of key 
observational properties requiring explanation.  The present shock acceleration models 
(Baring 2004) predict a wide range for the power law index of the particle energy 
distribution ($1.5\la p_1\la 3$), accommodating the observed range of the photon indices 
$1.2\la \alpha\la 2$.  We note that as an alternative explanation,  the particle distribution
described by a broken power law can reproduce
the observed hard spectrum in the context of the synchrotron emission
model provided that the particle distribution below the break energy  
is very steep.  In the synchrotoron spectrum, the corresponding photon energy 
of the break is $E_1=3he\Gamma_1B\sin\theta_p$, 
 where  $\Gamma_1$ is Lorentz  factor at the break in the particle disctribution and 
$\theta_p$ is the pitch angle.  For a very hard particle distribution below $\Gamma_1$,
the photon index changes from $\alpha=(p_1+1)/2$ (in the slow cooling
regime) above energy $E_1$ to $\alpha=2/3$ below $E_1$, 
which is the photon index in the spectral tail of the synchrotron radiation from a single 
particle.  Therefore, if the typical photon energy $E_1$ is larger than 10~keV, the 
spectrum in the 1-10~keV energy band can be characterized by a photon index less 
than 1.5.  This interpretation may be supported by the recent SUZAKU observations, which 
found evidence of spectral curvature and, in particular, a spectral break around 
4.5~keV (Uchiyama et al. 2009).

Accordingly, in this study, we seek to place constraints on the properties of the 
pulsar wind of PSR B1259-63 by developing an emission model that makes use of 
a stellar wind model for Be stars composed of a polar wind and an equatorial disk wind. 
The solution for the distribution of the particles of the shocked flow accounts for  
both synchrotron and inverse Compton losses, and the neglected 
corrections of the Klein-Nishina scattering cross section are included. 
In addition to the emission from the shocked pulsar wind, we also discuss the high 
energy emission from the inverse Compton process operating on the unshocked wind, 
which scatters the optical photons from the Be star. The numerical results describing 
the temporal behavior of the emergent flux and power law index in the X-ray and 
$\gamma$-ray energy band over the orbit of the system  are presented in \S 3. Finally, we 
discuss the results and future work in the last section.
 
\section{Theoretical model}
\label{model}

\subsection{Stellar wind and shock distance}

The dynamical pressure balance between the pulsar wind and the stellar outflow 
determines the location of the shock, that is, 
\begin{equation}
\frac{\dot{E}_{sp}}{4\pi cr_s^2}=\frac{\dot{M}v_w(R_s)}{4\pi f_{\Omega}R_s^2}
\left(\frac{v_{re}}{v_w}\right)^2, 
\label{balance}
\end{equation}
where $R_s=d-r_s$ is the distance of the shock from the stellar 
surface and $d$ is the orbital separation between the pulsar and the star. Here, we 
make use of the fact that the orbital separation $\sim 0.7-10$~AU and the shock 
distance $\sim 0.2-3$~AU from the pulsar are much larger than the 
stellar radius, $R_* \sim 0.05$~AU. In addition, $\dot{M}$ and 
$f_{\Omega}$ are the mass loss rate from the Be star and its outflow fraction in 
units of $4\pi$~sr, 
respectively. The velocities $v_{re}$ and $v_w$ are the outflow velocity 
of the wind relative to the pulsar and its intrinsic outflow velocity.  
The outflow from the Be star is described by a fast low density 
polar wind and a slow dense equatorial disk wind
 (Waters 1986; Waters et al. 1988). 

For the polar wind, the mass loss rate of typical early Be stars is $\sim 10^{-9}$ 
to $10^{-8} M_{\odot}$ yr$^{-1}$ (Snow 1981).  We adopt  the velocity field of the 
polar wind to be that for a radiatively driven wind given by
\begin{equation}
v_w(r)=v_0+(v_{\infty}-v_0)(1-R_*/R),
\end{equation}
where $v_0(\sim 20~\mathrm{km})$ is the initial velocity 
and $v_{\infty}$ is the terminal velocity, which is $\sim 2000$~km s$^{-1}$ 
for a typical Be star. Because the shock distance from the Be star is much larger 
than its 
stellar radius, we adopt $v_w(R_s)\sim v_{\infty}$ at the shock. 
Hence, the dynamical 
pressure balance condition (\ref{balance}) leads to a shock distance  
\begin{equation}
r_s=\frac{(\dot{E}_{sp}/cv_{\infty}\dot{M}_{p}/f_{\Omega,p})^{1/2}
}{1+(\dot{E}_{sp}/cv_{\infty}\dot{M}_p/f_{\Omega,p})^{1/2}}d,
\label{pshock}
\end{equation}
where we use the condition that $v_{re}\sim v_{\infty}$ because the terminal wind  
velocity is much greater than the typical orbital velocity of the pulsar, which is 
$v_{k}\sim(GM_*/a)^{1/2}\sim 40$~km s$^{-1}$ for a stellar mass corresponding to 
$M_*\sim 10M_{\odot}$ and semi major axis of $a\sim 5$ AU. 

For the dense equatorial flow, the typical mass loss rates are of the order of 
$\dot{M}_{e}=10^{-7}M_{\odot}$ yr$^{-1}$ (Waters 1986), and the velocity distribution 
is described by 
\begin{equation}
v_{w}(R)=v_0(R/R_*)^m,
\label{ewind}
\end{equation}
where the index $m$ is in the range of $0<m<2$, depending on the Be star
(Lamers \& Waters 1987).  In this paper, we adopt $m=0.4$, which is indicated 
for some  Be stars (Waters 1986).  

To determine the orbital phases when the pulsar wind interacts with the equatorial
disk of the Be star, we make use of the radio observations. Specifically, the non-pulsed 
radio emission, which is likely related to the interaction between the pulsar and the 
disk (Ball et al. 1999), appears between $80^{\circ}\la\theta\la 300^{\circ}$
 where $\theta=180^{\circ}+\phi$ with $\phi$ corresponding to the true anomaly. 
On the other hand, the pulsed radio emission from 
the pulsar magnetosphere disappears in the range $80^{\circ}\la\theta\la 280^{\circ}$.  
Since the eclipse
 is caused by 
absorption in the dense wind from the Be star (Johnston et al.  1992),
 we assume that 
the pulsar wind interacts with the thick equatorial disk-like outflow in the 
orbital phase from $80^{o}\la\theta\la 300^{o}$. 

\begin{figure}
\epsscale{.80}
\plotone{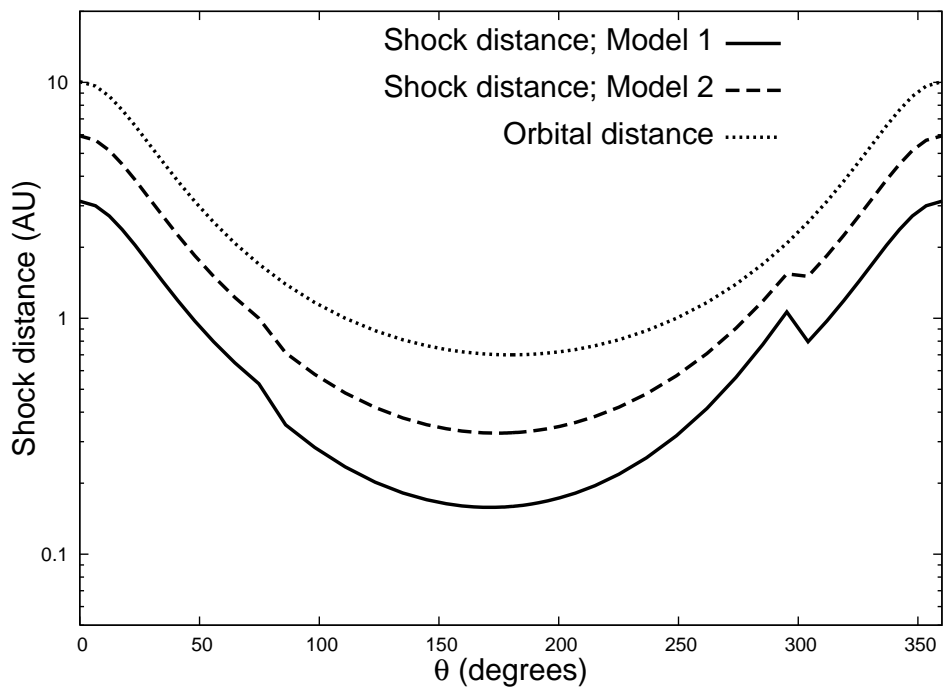}
\caption{Calculated  shock radius as a function of the orbital phase 
($\theta=\phi+\pi$). Model~1 (solid line);  
$\dot{M}_p/f_{\Omega}=10^{-8}M_{\odot}$~/yr 
and  $\dot{M}_e/f_{\Omega,e}=10^{-7}M_{\odot}$~/yr. 
 Model~2 (dashed line); $\dot{M}_p/f_{\Omega,p}=10^{-9}M_{\odot}$~/yr and
  $\dot{M}_e/f_{\Omega}=10^{-8}M_{\odot}$~/yr. The dotted line shows 
the orbital distance. }
\label{shockrad}
\end{figure}

Figure~\ref{shockrad} shows the distance to the shock front from the pulsar. 
The periastron is taken to be at $\theta=180^{\circ}$.
The solid and dashed lines are results for  
(Model 1) $\dot M_p/f_{\Omega,p}=10^{-8}M_{\odot}$/yr and $\dot M_e/f_{\Omega,e}
=10^{-7}M_{\odot}$/yr and (Model 2) $\dot M_p/f_{\Omega,p}=10^{-9}M_{\odot}$/yr
 and  $\dot M_e/f_{\Omega,e}=10^{-8}M_{\odot}$/yr, respectively. 
In the solid and dashed line, the discontinuities 
at $\theta=80^{\circ}$ and $300^{\circ}$  
reflect our assumption that the stellar wind discontinuously changes between 
the polar and  equatorial winds at those orbital phases.

\subsection{Dynamics of unshocked flow}
\label{dyunshock}

We assume the pulsar wind to be isotropic and established beyond the light 
cylinder.  The magnetization parameter at the light cylinder, $\sigma_L$, takes the 
value  
\begin{equation}
\sigma_L=\frac{B_{L}^2}{4\pi \Gamma_L n_L m_ec^2}
\sim 3\times 10^3\left(\frac{B_L}{10^{5}\mathrm{G}}\right)
\left(\frac{R_{lc}}{10^{8}\mathrm{cm}}\right)
\left(\frac{\kappa}{10^{3}}\right)^{-1}
\left(\frac{\Gamma_L}{10^{3}}\right)^{-1}
\end{equation} 
in which we used $n_{L}=\kappa n_{GJ}$ where $n_{GJ}$ is the Goldreich-Julian 
number density at the light cylinder, $\kappa$ is the multiplicity, and $R_{lc}$ 
is the radius of the light cylinder.  In pulsar models, most of the 
plasma is ejected from 
the light cylinder with a Lorentz factor $\Gamma_L\sim 10^{2-3}$ and $\kappa\sim 10^{2-3}$ 
(Daugherty \& Harding 1996; Hibschman \& Arons 2001), implying a magnetically 
dominant flow $\sigma_L\gg 1$. Between 
the light cylinder and the terminal shock, the magnetic field may 
be dissipated  so that the magnetic energy is converted to the particle energy 
of the flow (Coroniti 1990; Lyubarsky \& Kirk 2001).  If the magnetic 
energy were completely converted into the particle energy, 
the energy conservation implies that the terminal Lorentz factor 
of the bulk flow is 
\begin{equation}
\Gamma_1\sim \Gamma_L\sigma_L\sim 3\times 
10^6\left(\frac{B_L}{10^{5}\mathrm{G}}\right)
\left(\frac{R_{lc}}{10^{8}\mathrm{cm}}\right)
\left(\frac{\kappa}{10^{3}}\right)^{-1}.
\label{termL}
\end{equation}
After complete energy conversion, most of the energy of the flow 
is  carried by the particles with the Lorentz factor of $\Gamma\sim \Gamma_1$. 
In this case, a mono-energetic distribution, $f(r,
\Gamma)= n_1(r_s/r)^2\delta(\Gamma_1(r)-\Gamma)$, or very hard distribution 
with a power law index (say, less than unity)
  between $\Gamma_{L}<\Gamma<\Gamma_1$ may be applied for the energy 
distribution of the unshocked particles even though the pulsar wind particles 
 are ejected  from the light cylinder with a 
 power law index larger than  unity  above $\Gamma_{L}$ (which is 
expected from the of the pair-cascade process above polar cap studied 
by  Hibschman \& Arons 2001).  This ultra-relativistic flow collides 
with the stellar wind to form a shock, where the kinetic energy of 
the pulsar  wind is converted to the internal energy  of the wind.
 At the shock, some electrons and positrons are further 
accelerated to very high energy above $\Gamma_1$. (\S~\ref{dysh})

The emission from the inverse Compton process of the unshocked wind can also 
contribute to the high energy emission from $\gamma$-ray binaries. 
For example, Ball \& Kirk (2000) and Ball \& Dodd (2001) computed 
 the spectra due to the inverse Compton process 
from the unshocked wind  with (i) a mono-energetic distribution of 
the electron distribution, (ii) a mono-energetic distribution of background 
photon fields.  
In this paper, we improve upon these earlier treatments by calculating spectra 
using the Planck distribution for the background photon field and using the 
differential Klein-Nishina cross section with an anisotropic photon field 
(also see Khangulyan et al. 2007 and Cerutti et al. 2008).
In addition, to determine the sensitivity of the resulting spectrum to 
the input particle energy distribution, we also consider a power law distribution 
for the unshocked particles.
The power per unit energy and per unit solid angle of the inverse Compton process is 
described in Takata \& Chang (2007) and reference therein as 
\begin{equation}
\frac{dP_{IC}}{d\Omega}=D^2\int_0^{\theta_c}(1-\beta\cos\theta_0)I_b/h
\frac{d\sigma'}{d\Omega'}d\Omega_0, 
\label{inver}
\end{equation}
where $D=\Gamma^{-1}(1-\beta\cos\theta_1)^{-1}$ with $\theta_1$ and $\theta_0$ 
describing the angle between the direction of the particle motion and the propagating 
direction of the scattered photons and background photons respectively. In addition, 
$I_b$ is the background photon field and $\theta_c=\sin^{-1}R_*/r$ expresses the 
angular size of the star as seen from the point $r$. Here,
$d\sigma'/d\Omega'$ is the differential Klein-Nishina cross section. 
Since the electrons and positrons of the unshocked wind lose only about ~1\% of their 
energy by interaction with the stellar photons, we can ignore the evolution of 
the distribution function in the unshocked particles.
 
\subsection{Dynamics of shocked flow}

\label{dysh}
Following  Tavani \& Arons (1997), it is assumed that the electrons and 
positrons are accelerated to ultra-relativistic energies at a shock front that
terminates the pulsar wind of PSR B1259-63. The ultra-relativistic particles injected 
downstream emit non-thermal radiation via the synchrotron process and the inverse 
Compton process, the latter involving the upscattering of the stellar optical photons 
of SS~2883 to the TeV energy band.  For the pulsar wind, spherical symmetry is assumed. 
The shock distance from the pulsar is also treated as spherically symmetric to the 
lowest order approximation.  Although the pulsar wind and stellar wind are taken to 
be isotropic, the geometry of the shock is anisotropic.  In particular, Bogovalov 
et al. (2008) investigated the collision between the isotropic pulsar wind and the 
stellar wind in the hydrodynamical limit, finding that the geometry of the shock
is a function of $\eta\equiv\dot{E}_{sp}/c\dot{M}v_w$.  In particular, they find 
that the collision of the two winds produces an open shock geometry when $\eta>1.25\times 
10^{-2}$.  Thus, the shock geometry for the PSR B1259-63/SS 2883 system may be a 
function of orbital phase due to the factor, $\eta$, reflecting the different 
properties of a fast polar wind from a slow dense equatorial wind. In this case, within our level of the approximation, the inferred magnetization parameters should be considered as spherically averaged values. 

For the post shock flow, the non-dimensional radial four velocity $u_2$, 
the proper number density $n_2$, the magnetic field $B_2$ and the gas pressure $P_2$ 
 at the shock are derived using the jump conditions of a perpendicular MHD shock. For 
the particle kinetic energy dominant flow, that is for the  low $\sigma$ regime
($\sigma \ll 1$), we obtain (Kennel \& Coroniti, 1984a, b)
\begin{equation}
u_2=\left(\frac{1+9\sigma}{8}\right)^{1/2}, 
\end{equation}
\begin{equation}
n_2=\frac{n_1u_1}{u_2},
\label{number2}
\end{equation}
\begin{equation}
B_2=3B_1(1-4\sigma),
\label{b2}
\end{equation}
and 

\begin{equation}
\frac{P_2}{n_2u_1m_ec^2}=\frac{1}{\sqrt{18}}(1-2\sigma).
\label{press2}
\end{equation}

For a steady state, specification of the velocity field  $u(r)$ of the downstream 
flow as a function of the radial distance from the shock front allows one to 
determine the variation of the  number density and the magnetic field as a function 
of the radial 
distance. From conservation of the number flux and magnetic flux, we obtain 
\begin{equation}
n(r)u(r)r^2=\textrm{constant}
\end{equation}
and 
\begin{equation}
u(r)B(r)r/\Gamma_w=\textrm{constant},
\end{equation}
where $\Gamma_w=\sqrt{1+u^2}$ is the Lorentz factor of the shocked wind.
In this study, we describe the velocity field with  $u(r)=\textrm{constant}$ 
along the flow, because (i) the high energy emission occurs in the vicinity 
of the shock, and (ii) recent numerical simulations in the hydrodynamical 
limit imply that the post shock bulk flow for the binary system does not simply
 decrease because of a rapid expansion of the flow  in the downstream region,
 unlike the case for an isolated pulsar (Bogovalov et al. 2008). 

To determine the distribution function of the shocked 
particles, we refer  the procedure adopted by Kennel \& Coroniti (1986a, b).
Because the shock acceleration theories indicate a break 
in the particle energy distribution in the post-shocked flow around 
the injection energy (e.g. Jones \& Ellison 1991; Ellison \& Double 2004), 
 we assume that the downstream particles form a broken power law distribution 
described by
\begin{equation}
f_2(\Gamma)=\frac{K_n}{4\pi}
\left\{
\begin{array}{@{\,}ll}
\left(\frac{\Gamma}{\Gamma_1}\right)^{-p_1} & (\Gamma_1<\Gamma\le\Gamma_{max})
\\
\left(\frac{\Gamma}{\Gamma_1}\right)^{-p_2} & (\Gamma_{min}<\Gamma\le\Gamma_1).
\end{array}
\right .
\label{power}
\end{equation} 
 The maximum Lorentz factor is determined from the condition
$\Gamma_{max}=\textrm{min}(\Gamma_{s}, \Gamma_{g})$, 
where $\Gamma_s$ is the Lorentz factor for which the particle acceleration 
time scale $t_{ac}\sim \Gamma_1 m_e c/eB_1$ is equal to the synchrotron 
radiation loss time scale, $\tau_{sync}\sim 9m_e^3c^5/e^4B^2_1\Gamma_1$. Here 
$\Gamma_g$ is the Lorentz factor for which the gyration radius 
is larger than the shock radius ($\sim r_s$).  For the minimum Lorentz factor 
 $\Gamma_{min}$, we adopt $\Gamma_{min}\sim\Gamma_{L}\sim10^{3}$, related to 
the Lorentz factor of the pairs produced in the magnetosphere. Our results are 
found to be insensitive to this choice.

The distribution function in equation~(\ref{power}) 
 is described by three quantities 
$K_n$, $p_1$ and $p_2$, and we require three conditions to specify the energy 
distribution, $f_2$.   However, the theoretical  
uncertainties in the detailed microphysics of the acceleration model preclude 
a determination of the power law index $p_1$ of the accelerated particles at the 
shock. Thus, we take the index $p_1$ as a model parameter ranging between 1.5 and 3.
Because the distribution function $f_2$ must satisfy 
the conditions $n_2=\int\int f_2(\gamma)d\Gamma d\Omega$ and
$\epsilon_p=m_ec^2\int\int \Gamma f_2(\Gamma)d\Gamma d\Omega$, where
$\epsilon_p$ is the energy density of the particles, we use these 
two conditions to calculate the two 
parameters, $K_n$ and $p_2$. For a relativistic gas,
 the energy density $\epsilon_p$ is related to the gas pressure $P_2$ as
$\epsilon_p=3P_2$.  On the other hand,
the quantities $n_2$ and $P_2$ are 
determined by the jump condition at the shock
 and are calculated  from equations (\ref{number2}) and (\ref{press2})
 in the low $\sigma$ regime ($\sigma\ll 1$).

Note that in Kennel \& Coroniti (1986a, b), a single power 
law distribution $f\propto \Gamma^{-p_1}$ between $\Gamma_{min}\le\Gamma\le
\Gamma_{max}$ is assumed, and the normalization factor $K_{n}$ and 
$\Gamma_{min}$ are calculated from the two conditions on $n_2$ and 
$\epsilon_p$. As we discussed above, however, we assumed a break in the 
distribution function at $\Gamma_1$ and, alternatively, calculate the 
normalization factor $K_n$ and the index $p_2$ below the Lorentz factor 
$\Gamma_1$, because (i) the shock acceleration theories indicate a break 
in the particle energy distribution in the post-shocked flow around 
$\Gamma_1$ and (ii) $\Gamma_{min}$ will be related to 
the Lorentz factor of the pairs produced in the magnetosphere.

In the downstream region, the particles 
lose their energy via adiabatic expansion 
and radiation losses. To describe a general behavior of the time evolution of 
the Lorentz factor of  the  particles under the isotropic  distribution,
 we integrate any angular dependences 
(e.g. the pitch angle and 
 collision angle for the inverse-Compton process) and obtain  
\begin{equation}
\frac{d\Gamma}{dt}=\frac{\Gamma}{3 n}\frac{d n}{dt}-\left(\frac{d\Gamma}{dt}
\right)_{syn}-\left(\frac{d\Gamma}{dt}\right)_{IC}, 
\label{levol}
\end{equation}
where the synchrotron energy loss rate is given as
\begin{equation}
\left(\frac{d\Gamma}{dt}\right)_{syn}=\frac{4e^4B^2\Gamma^2}{9m_e^3c^5}, 
\label{syloss}
\end{equation}
and the inverse Compton energy loss rate is  
\begin{equation}
\left(\frac{d\Gamma}{dt}\right)_{IC}=\int\int(E-E_s)
\frac{\sigma_{IC} c}{m_ec^2 E_s}\frac{dN_s}{dE_s}dE_sdE, 
\label{icloss}
\end{equation}
where $dN_s/dE_s$ is the stellar photon field distribution from the Be star, 
and the cross section $\sigma_{IC}$, which for $\Gamma\gg 1$, is described by 
\begin{equation}
\sigma_{IC}=\frac{3\sigma_T}{4\Gamma^2}
\left[2q\textrm{ln}q +(1+2q)(1-q)+\frac{(\Gamma_q q)^2(1-q)}{2(1+\Gamma_q q)}
\right]
\label{ic}
\end{equation}
where $\Gamma_q=4\Gamma E_s/m_ec^2$, $q=E_0/\Gamma_q(1-E_0)$ 
with $E_0=E/\Gamma m_ec^2$ and $1/(4\Gamma^2)<q<1$ (Blumenthal \& Gould 1970). 
With equation (\ref{levol}), the Lorentz factor is a function of 
the initial Lorentz factor $\Gamma_2$ at the shock and the distance from the 
shock. From the number conservation in the phase 
space, the distribution function at the distance $r$ can be calculated from 
\begin{equation}
f(r,\Gamma)=\frac{n}{n_2}f_2 \frac{d\Gamma_2}{d\Gamma}.
\end{equation}

\begin{figure}
\epsscale{.80}
\plotone{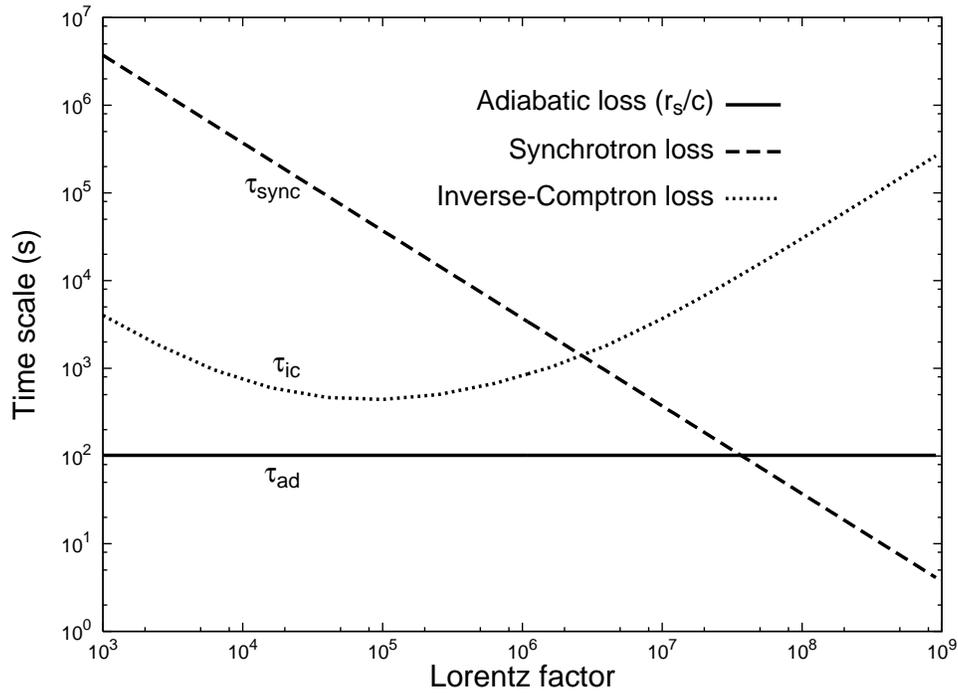}
\caption{The time scale of the adiabatic loss (solid-line), the synchrotron 
loss (dashed-line) and the inverse-Compton loss (dotted-line) at 
the periastron as a function of the Lorentz factor. The results are 
for $\sigma=10^{-2}$ with $\dot{M}_p/f_{\Omega}=5\cdot 10^{-9}M_{\odot}$~/yr 
and  $\dot{M}_e/f_{\Omega,e}=5\cdot 10^{-8}M_{\odot}$~/yr.}
\label{loss}
\end{figure}

\begin{figure}
\epsscale{.80}
\plotone{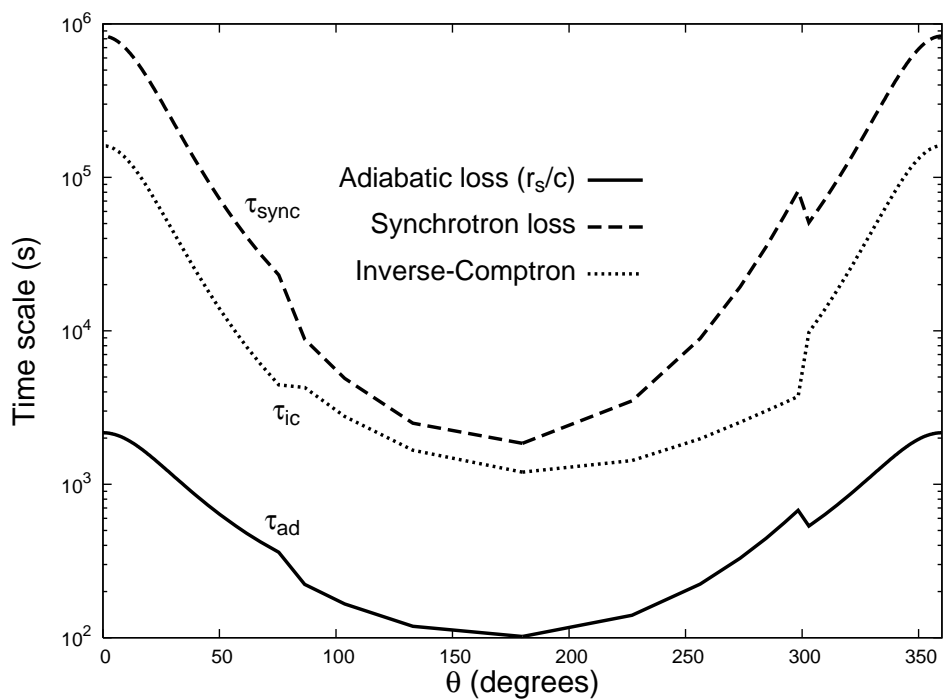}
\caption{The time scales of the energy loss for the particles with 
 the Lorentz factor $2\times10^6$ as a function of the orbital phase.  
The results are for  $\sigma=10^{-2}$ with $\dot{M}_p/
f_{\Omega}=5\cdot 10^{-9}M_{\odot}$~/yr 
and  $\dot{M}_e/f_{\Omega,e}=5\cdot 10^{-8}M_{\odot}$~/yr.}
\label{tscale}
\end{figure}

Figure~\ref{loss} compares 
the typical time scale for energy loss at the periastron  due to 
 adiabatic expansion (solid-line), 
synchrotron radiation (dashed-line) and inverse-Compton (dotted-line) 
processes as a function of the Lorentz factor. The results are for 
 $\sigma=10^{-2}$ with $\dot{M}_p/
f_{\Omega}=5\cdot 10^{-9}M_{\odot}$~/yr 
and  $\dot{M}_e/f_{\Omega,e}=5\cdot 10^{-8}M_{\odot}$~/yr. 
We find that the adiabatic loss dominates the radiation 
energy losses, except for the very high energy regime ($\Gamma>10^8$), 
for which the synchrotron loss is more significant  than the adiabatic loss 
(see also Khangulyan et al. 2008). 
The adiabatic loss dominates throughout the orbital phase, as indicated in 
Figure~\ref{tscale}, which compares the various cooling time scales of the particles 
for a Lorentz 
factor $\Gamma=2\times 10^6$ as a function of the orbital phase. In the case that 
the adiabatic loss describes the evolution of the Lorentz factor of the shocked particles,  
the distribution function evolves as $f(\Gamma)\propto r_s/r$.

\begin{figure}
\epsscale{.80}
\plotone{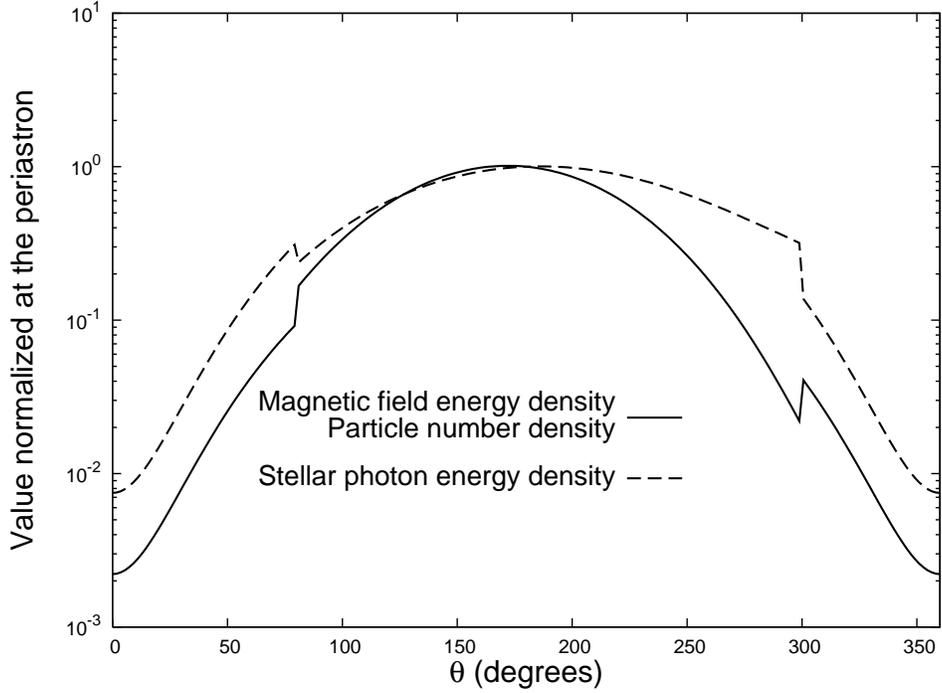}
\caption{Variation of the magnetic field energy and  particle number density 
(solid line), which evolve as $\propto 1/r_s^2$, and the stellar photon energy 
density (dashed line) at the shock.  The curves of the magnetic field density 
and the particles number density, which are proportional to the inverse square 
of the shock distance $\propto r^2_s$,  overlap in the figure. 
 The values are normalized at the periastron, where the magnetic field energy,
 the particle number density and the stellar photon energy density 
are $\sim 0.018~\mathrm{erg/cm^3}$,$\sim0.37~\mathrm{/cm^3}$ and 
 $\sim 34~\mathrm{erg/cm^3}$, respectively. 
  The results are for  
$\sigma=10^{-2}$ with $\dot{M}_p/
f_{\Omega}=5\cdot 10^{-9}M_{\odot}$~/yr 
and  $\dot{M}_e/f_{\Omega,e}=5\cdot 10^{-8}M_{\odot}$~/yr.}
\label{energy}
\end{figure}

For the inverse-Compton process described by the full functional 
form of the Klein-Nishina cross section, the cooling time scale is proportional to the 
Lorentz factor in the Klein-Nishina regime ($\Gamma\geq 10^5$), while it is proportional 
to the inverse of the Lorentz factor in the Thomson regime ($\Gamma\leq 10^5$). 
The suppression of the inverse-Compton process in the Klein-Nishina regime causes
the cooling time scale  of the inverse-Compton process to be 
much longer  than that of the  synchrotron process, although the 
energy density of the stellar photons 
is much larger than that of the magnetic 
field as Figure~\ref{energy} indicates. 
We note that the 1-10~keV photons are emitted 
by the synchrotron radiation process from particles with a Lorentz factor 
of $\Gamma\sim 10^6-10^7$ (depending on the orbital phase),
 indicating the particles are in slow 
radiation cooling regimes as Figures~\ref{loss} and 
\ref{tscale} show.

It should be noted that the mass loss rate affects the cooling time scale 
of the particles since the position of the shock is a function of this loss 
rate. To determine the relative importance of the various processes, we recast
the cooling time scales as a function of the shock distance from the pulsar. 
In this form, the adiabatic loss time scale $\tau_{ad}\propto 1/r_s$, the 
synchrotron loss time scale $\tau_{sync}\propto 1/r_s^{3/2}$, and the 
inverse-Compton loss time scale $\tau_{ic}\propto (d-r_s)^2$, where  
the relation that Lorentz factor of the emitting particles is proportional 
to $\Gamma\propto 1/B^{1/2}$ has been used for a specific energy of the 
synchrotron photons. Because the synchrotron radiation loss time scale is more 
sensitive to the shock distance than the adiabatic loss time scale, the 
synchrotron loss dominates others provided that the shock is located within 
the critical distance from the pulsar.  For the X-ray emitting particles, 
however, one finds that the synchrotron loss dominates with  the shock distance 
smaller than $r_s <2\times 10^{-4}$~AU, providing  the mass loss rate is 
larger  than $\dot{M}>0.1M_{\odot}$~/yr from the pulsar,
 which is not expected the typical 
mass loss rate of the Be star. On 
the other hand, the inverse-Compton loss may dominate at the periastron 
passage because the shock lies close to the Be star.  For the X-ray emitting 
particles, inverse-Compton losses are more important if the shock is located 
at a distance smaller than $R_s=d-r_s<0.2$~AU from the Be star.  The mass 
loss rate for a shock distance $R_s\sim 0.2$~AU at the periastron is $\dot{M}
\sim 2\times 10^{-9}M_{\odot}$~/yr, which is lower than the typical mass loss 
rate of the equatorial wind, but in the range of the polar wind.  In the 
present study, with our assumption that the pulsar wind mainly interacts with 
the equatorial wind at the periastron passage, the adiabatic loss processes 
for X-ray emitting particles dominate for the typical mass loss rates of Be 
stars.

In Tavani \& Arons (1997), the inverse-Compton cooling time
 is proportional to the inverse of the 
Lorentz factor for the inverse-Compton process in the Klein-Nishina regime 
 because they adopted an approximate formula for the energy loss rate of 
the inverse Compton process as $\dot{\Gamma}_{IC}=\alpha_{IC}\dot{\Gamma}_{T}$,
 where 
$\dot{\Gamma}_T=c\sigma_TB\Gamma^2/6\pi m_ec^2$ is the energy loss
 rate in the Thomson limit and $\alpha_{IC}$ is a proportionality factor. 
For the form used by Tavani \& Arons (1997), the inverse-Compton energy loss 
dominates the adiabatic loss near the periastron, and 
the strong inverse Compton process 
softens the particle distribution. 
 However, we found in Figure~\ref{loss} and ~\ref{tscale} 
 that with the full 
formula of (\ref{ic}) of the cross section of the inverse-Compton process, the 
adiabatic loss is greater than the inverse-Compton loss for entire orbit. 
 Therefore, the spectral fitting obtained in  Tavani \& Arons (1997) 
is likely to be modified using the full formula for the cross section of the 
inverse Compton process.

The synchrotron power per unit energy emitted by each electron is 
calculated from Rybicki \& Lightman (1979) as 
\begin{equation}
P_{syn}(E)=\frac{\sqrt{3}e^3B\sin\theta_{p}}{hm_ec^2}F
\left(\frac{E}{E_c}\right),
\label{synchro}
 \end{equation}
where $\theta_p$ is the pitch angle, $E_c=3he\Gamma^2 B\sin\theta_p/
4\pi m_e c$ 
is typical photon energy and  
$F(x)=x \int_x^{\infty} K_{5/3} (y)dy$, where $K_{5/3}$ is the modified 
Bessel function of order 5/3.
For the pitch angle, we use the averaged value corresponding to 
$\sin^2\theta_p=2/3$. 
The power per unit energy and per unit solid angle of the inverse Compton process is 
described by equation (\ref{inver}).

\begin{figure}
\epsscale{.80}
\plotone{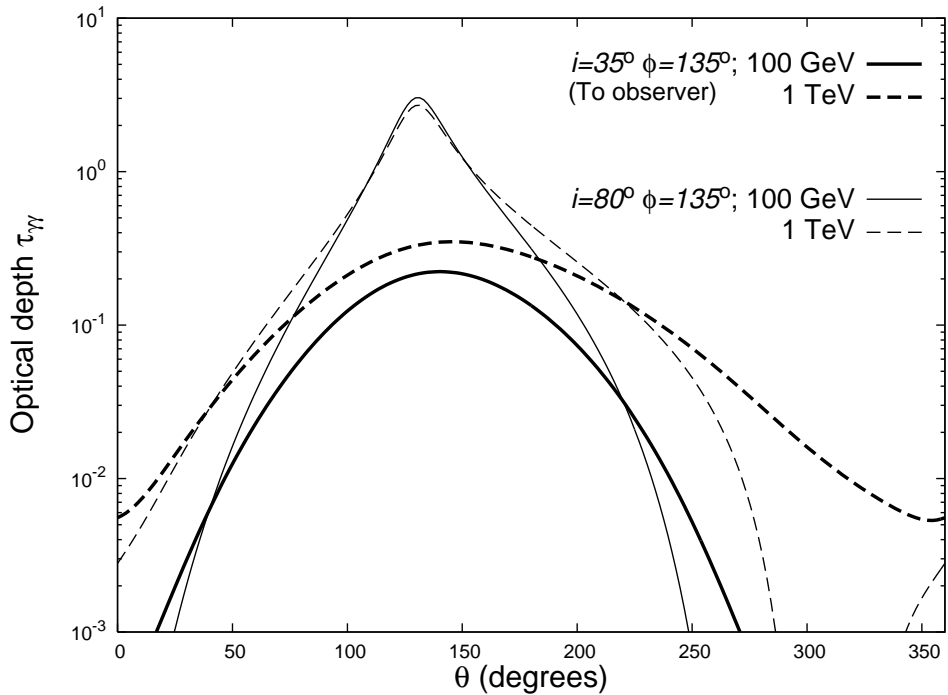}
\caption{Optical depth associated with the creation of pairs by photons 
with energy 100~GeV (solid-lines) and 1~TeV (dashed-lines). The thick and thin 
lines represent the results for  the photons propagating 
to  the observer  $(i,~\phi)=(35^{\circ},~ 130^{\circ})$  and  
the direction with $(80^{\circ},~ 130^{\circ})$, respectively.
}
 \label{depth}
\end{figure}

For B1259-63/SS2883, the viewing geometry was determined by the observations, 
where the inclination angle of the plane of the orbit with respect to the sky 
is $i\sim 35^{\circ}$ and the true anomaly of the direction of Earth is about 
$\phi\sim 130^{\circ}$ (Johnston et al. 1996). The model spectrum of the 
emission from the shocked wind measured on Earth is calculated from 
\begin{equation}
EF_E=\frac{Ee^{-\tau_{\gamma\gamma}}}{d^2}
\int_{r_s}^{r_{max}}dr\int d\Gamma
r^2f(\Gamma)\left(P_{syn}+\int\frac{dP_{IC}}{d\Omega}d\Omega
\right),
\end{equation}
where $\tau_{\gamma\gamma}$ is the optical depth for the pair-creation process.
 The optical depth of the photons with 100~GeV and 1~TeV 
 propagating to the observer as 
a function of the orbital phase is illustrated in Figure~\ref{depth} with 
the thick solid and dashed lines, respectively. We find from figure that 
the optical depth is smaller than unity, indicating the absorption of the 
photons propagating to the observed does not modify the level of the flux.  
The main reason of the optical depth smaller than unity 
is the viewing geometry of the observer. For example,  
 most of the photons propagating to direction 
$i=80^{\circ}$ and $\phi=135^{\circ}$ are absorbed 
by the stellar photons near the periastron, where the optical depth is greater 
than unity  as indicated by thin lines in Figure~\ref{depth}. 
 We integrate  the emission from the shock to the distance
$r_{max}=2r_s$, following the result obtained by Bogovalov et
al. (2008) that the emitting region, which lies between the shock and
the contact discontinuity, is confined near the shock.

\section{Results}
\subsection{Radiation spectrum from a shocked wind}
\begin{figure}
\epsscale{.80}
\plotone{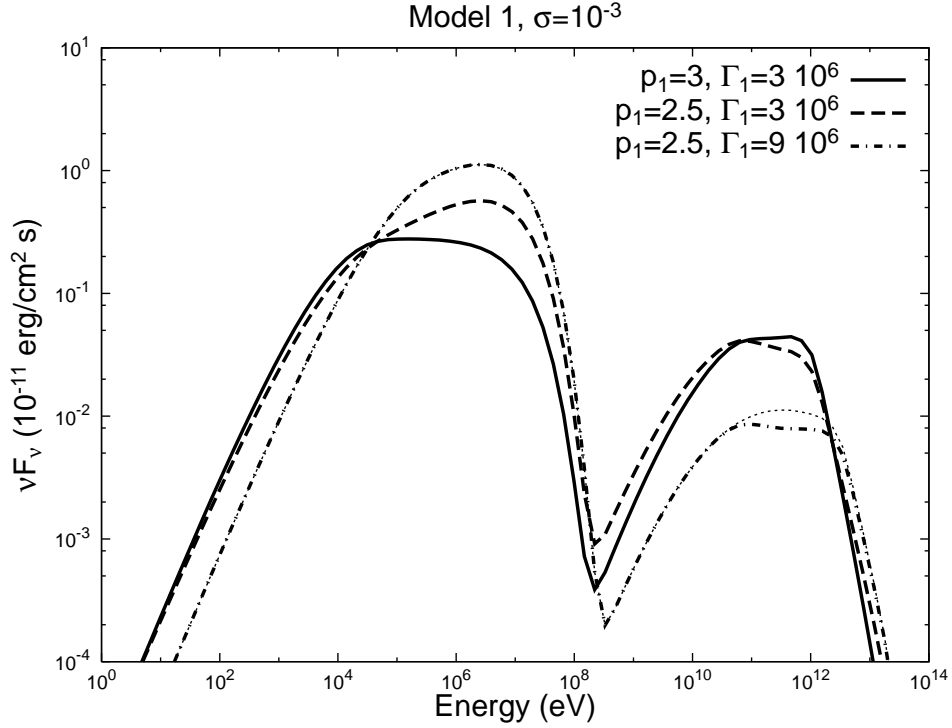}
\caption{The emission spectra as a function of energy from the shocked 
particles at periastron. All results are for $\sigma=10^{-3}$ and for a shock 
distance calculated with Model~1. The  photon index $p_1$ and the Lorentz factor 
$\Gamma_1$ are $(p_1,~\Gamma_1)=(3,~3\times 10^6)$ for the solid line, $(2.5,
~3\times 10^6)$ for the dashed line, and $(2.5,~9\times 10^6)$ for the 
dashed-dotted line.  The thin dotted line is the result for $(p_1,~\Gamma_1)=
(2.5,~9\times 10^6)$ ignoring the effect of the pair-creation process. }
 \label{general}
\end{figure}

The calculated spectra from the optical to very high-energy bands at 
the periastron due to the shocked wind for $\sigma=10^{-3}$ and for a shock 
distance determined with Model~1 is illustrated in Figure~\ref{general}. 
The different line types represent the results for the different photon index 
$p_1$ and the Lorentz factor $\Gamma_1$.  Specifically, the solid line, dashed 
line, and dash dotted line correspond to $(p_1,~\Gamma_1)=(3,~3\times 10^6)$,
$(2.5,~3\times 10^6)$, and $(2.5,~9\times 10^6)$ respectively.  The thin-dotted 
line is result for $(p_1,~\Gamma_1)= (2.5,~9\times 10^6)$ with the 
pair-creation process neglected. 

The calculated spectra below 100~MeV is due to synchrotron radiation, and it can 
be seen (see Figure~\ref{general}) that the photon index of the spectrum changes at two 
energies around 10-100~keV and 10~MeV. The lower ($E_1$) and higher ($E_{max}$) energy 
spectral breaks correspond to the typical energies of the synchrotron radiation obtained 
with  the Lorentz factor $\Gamma_1$ and $\Gamma_{max}$, respectively.  The solid 
and dashed-dotted lines reveal that the energy of the lower spectral break ($E_1$) 
increases from 10~keV to 100~keV as the Lorentz factor $\Gamma_1$ increases from 
$\Gamma_1=3\times 10^6$ to $\Gamma_1=9\times 10^6$, reflecting the fact that the energy 
of the synchrotron photons is proportional to the square of the Lorentz factor $\Gamma^2$.

Between the energies $E_1$ and $E_{max}$, the emission of synchrotron radiation 
is in the slow cooling regime (section~\ref{dysh}), 
described by a photon index given by $\alpha=(p_1+1)/2$. 
For example, the photon index of the solid line in Figure~\ref{general} is about 
$\alpha\sim 2$, reflecting the power law index $p_1=3$ of the electron distribution.
Upon comparison of the results for the three cases in Figure~\ref{general}, 
we find that the flux increases with
 (i) decreasing values of the power law index 
$p_1$ of the electron distribution as indicated by the solid 
and dashed lines,  
and (ii) increasing values of the Lorentz factor $\Gamma_1$ as indicated 
by the dashed and dashed-dotted lines.  The former result is a consequence 
of the fact that more particles are distributed at higher energies for a 
smaller power law index. For the latter case, the energy distribution of 
the particles becomes harder as $\Gamma_1$ increases, and the number of 
particles per unit Lorentz factor between $\Gamma_1<\Gamma<\Gamma_{max}$ 
increases with the Lorentz factor $\Gamma_1$. Therefore, the flux due to  
synchrotron radiation between $E_1<E<E_{max}$ increases with the Lorentz 
factor $\Gamma_1$. On the other hand, the flux below $E_{1}$ decreases 
with increasing values of the Lorentz factor $\Gamma_1$ because the particles 
per unit Lorentz factor below $\Gamma_1$ decrease as the particle 
distribution becomes harder with increasing Lorentz factor $\Gamma_1$. 
Below $E_1$ the spectral index changes from $\alpha=(p_1+1)/2$ to a very 
hard photon index $\alpha=(p_2+1)/2$ or $\alpha=2/3$ if $p_2\le 1/3$.
 
The inverse Compton process produces the emission above 100~MeV in the 
spectra of Figure~\ref{general}. With the typical energy of 
the stellar photons of $kT\sim$~2eV, the peak 
in the spectral energy distribution is located at $(m_ec^2)^2/kT\sim 0.1~$TeV 
as shown in Figure~\ref{general}. Upon comparing the dashed-dotted and 
thin-dotted lines, we find the flux level from the inverse Compton 
process including the effect of the absorption process does not differ 
significantly from the level calculated without its inclusion.  This is 
because the optical depth of the photons propagating toward  the observer is 
smaller than unity for the entire orbit, as the thick-lines in 
Figure~\ref{depth} reveal.
 This result is consistent with that obtained by Kirk et al. (1999) and Dubus 
(2006a).  As shown in the dashed-dotted line, however, we can see the effects 
of absorption on the shape of the spectrum due to the inverse Compton
 process.  In particular, the spectral shape with the absorption
 is characterized by 
a plateau between the spectral breaks at $\sim$50~GeV and $\sim$5~TeV.
This is because below 50~GeV and 5~TeV, 
the pair creation process is ineffective.

Recently, Sierpowska-Bartosik \& Bednarek (2008) argued that most of the 
inverse Compton photons which propagate close to the stellar surface are 
absorbed by the stellar photons, producing electron and position pairs. 
Provided that the shock position is relatively close to the Be star, 
these pairs emit additional very high energy photons associated with
 the inverse Compton process, 
resulting in a pair-creation cascade. 
As indicated by thin-lines in 
Figure~\ref{depth} for example, the optical depth of TeV photons 
emitted at a orbital phase $\theta\sim 120^{\circ}$ and toward
 $i=80^{\circ}$ and $\phi=135^{\circ}$ is larger than unity,
 implying most of photons are absorbed by the stellar photons 
and create the secondary pairs.
  Because the pairs gyrate around the 
magnetic field lines of the Be star with a pitch angle,
 the pairs can emit very 
high energy photons in directions differing from the propagation
 of the primary $\gamma$-rays. 
This implies that the higher-order generated pairs may emit the 
$\gamma$-ray radiation toward the observer even though the primary $\gamma$-rays 
are not directed toward the observer. As a consequence, Sierpowska-Bartosik \& Bednarek 
(2008) found a non negligible contribution from the pairs to the observed emission 
spectrum at epochs close to the periastron for PSR B1259-63/SS2883 system.  Incorporating  
this cascade process to our model would lead to additional emission between the 50~GeV 
and 5~TeV energy band since the $\gamma$-ray absorption by the stellar photons in these 
energy bands is more efficient. In this case, the plateau shape of the spectra seen in 
Figure~\ref{general} will be modified. Although the high-order generated pairs
will also emit  X-ray photons via the synchrotron radiation, which may contribute 
to the emission observed at the periastron passage, we have neglected   
these effects in our computations.

The total emission of the inverse Compton process decreases with an increase of the 
Lorentz factor $\Gamma_1$ (see the dashed and dashed-dotted lines in 
Figure~\ref{general}) as the scattering process is more efficient in the Thomson 
regime as compared with the Klein-Nishina regime. 
For the PSR B1259-64/SS2883 system, particles with 
Lorentz factors smaller than $m_ec^2/kT\sim 2.5\times 10^5$, 
scatter stellar photons in the Thomson regime. This indicates that the 
particles in the tail ($\Gamma<\Gamma_1$) of the energy distribution mainly 
contribute to the total emission of the inverse Compton process. Because 
the particle distribution becomes harder as the Lorentz factor $\Gamma_1$ 
increases, the number of particles per unit energy in the tail of the 
distribution decreases for a given observed energy, and as a result the 
total emission decreases.

\subsection{Flux in the 1-10~keV energy band}
\begin{figure}
\epsscale{.80}
\plotone{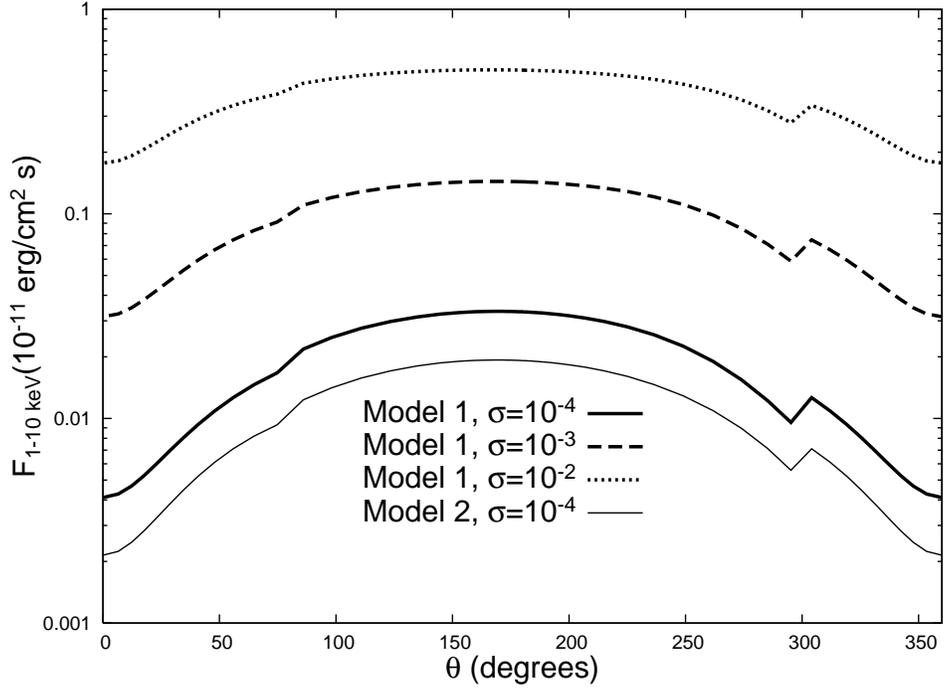}
\caption{Variation of the integrated flux of the synchrotron radiation
in the 1-10~keV energy band with respect to   orbital 
phase.  The thick-solid, dashed and dotted lines represents the flux 
 for $\sigma=10^{-4}$, 
$10^{-3}$ and $10^{-2}$ with $\dot{M}_p/f_{\Omega,p}=10^{-8}M_{\odot}$~/yr
and  $\dot{M}_e/f_{\Omega,e}=10^{-7}M_{\odot}$~/yr. The thin-dotted line 
is result obtained with   $\sigma=10^{-4}$, 
$\dot{M}_p=10^{-9}M_{\odot}/f_{\Omega,p}$~/yr
and  $\dot{M}_e/f_{\Omega,p}=10^{-8}M_{\odot}$~/yr.  
All results are for $\Gamma_1=3\times 10^{6}$ and $p_1=2.5$.}
 
\label{synflux}
\end{figure}

Figure~\ref{synflux} shows the calculated flux  in the 1-10 keV energy band
as a function of orbital phase.  The thick-solid, 
dashed and dotted lines show the results for $\sigma$ parameters of 
$10^{-4}$, $10^{-3}$ and $10^{-2}$ with  the shock distance calculated 
with $\dot M_p/f_{\Omega,p}=10^{-8}M_{\odot}$/yr and $\dot M_e/f_{\Omega,e}
=10^{-7}M_{\odot}$/yr. The thin-solid line shows 
the result for $\sigma=10^{-4}$ with the shock distance calculated 
with $\dot M_p/f_{\Omega,p}=10^{-9}M_{\odot}$/yr and $\dot M_e/f_{\Omega,e}
=10^{-8}M_{\odot}$/yr. We assume $\Gamma_1=3\times 10^6$ as the 
Lorentz factor of bulk motion of the unshocked wind and 
$p_1=2.5$ as the power law index of the distribution of   
the accelerated particles above $\Gamma>\Gamma_1$. We obtained the 
power law index $p_2\sim 0.7$ for  $\Gamma<\Gamma_1$ with the method 
described in section~\ref{model}.
 
 By comparing  the thick-solid, dashed and 
dotted lines, we find that the flux increases 
with the $\sigma$ parameter at a given orbital phase (i.e. the same shock distance). 
This is simply because the magnetic field at the shock calculated from equations 
(\ref{b1}) and (\ref{b2}) increases with $\sigma$, thereby increasing the emissivity 
associated with synchrotron radiation. 
Moreover, the calculated flux in the 1-10 keV energy band tends to increase 
as the pulsar moves from  the apastron to the periastron, that is, 
the luminosity is proportional to $L_{syn}\propto 1/r_s$. 
 The luminosity is approximately described by   $
L_{syn}\sim Vn_2P_s$, 
where $V$ is the volume of the emission region, $n_2$ is the number density 
of the particles and $P_s$ is power of synchrotron emission of a single
 particle. The volume
 of the emission region $V$ is  $V\sim r_s^3$, if the size of the system 
is less than the cooling length of 
the synchrotron emission. 
 The particle number density at the shock calculated from equation
(\ref{number2}) is proportional to $n_2\propto r_s^{-2}$, and the power 
of the synchrotron radiation is proportional to $P_s\propto B_2^2\propto 
r^{-2}_s$. As a result, the synchrotron luminosity is a function 
of the shock distance ($L_{syn}\propto 1/r_s$) and 
increases with decreasing radial distance of the shock from the pulsar. 

A comparison of  the thick- and thin-solid lines in Figure~\ref{synflux} 
describes the dependence of the synchrotron flux on the stellar wind models.
Model~1 assumes a denser stellar wind model than Model~2. 
As Figure~\ref{shockrad} illustrates  the shock distance from the pulsar 
for Model~1 is about a factor of 2 smaller than that for Model~2. 
Since the synchrotron luminosity is related to the shock distance 
as $L_{syn}\propto 1/r_s$, the  synchrotron radiation for Model~1
is about factor 2 greater than in Model~2, as shown in Figure~\ref{synflux}. 

\subsection{Photon index in the 1-10~keV energy range}
\label{index}
\begin{figure}
\epsscale{.80}
\plotone{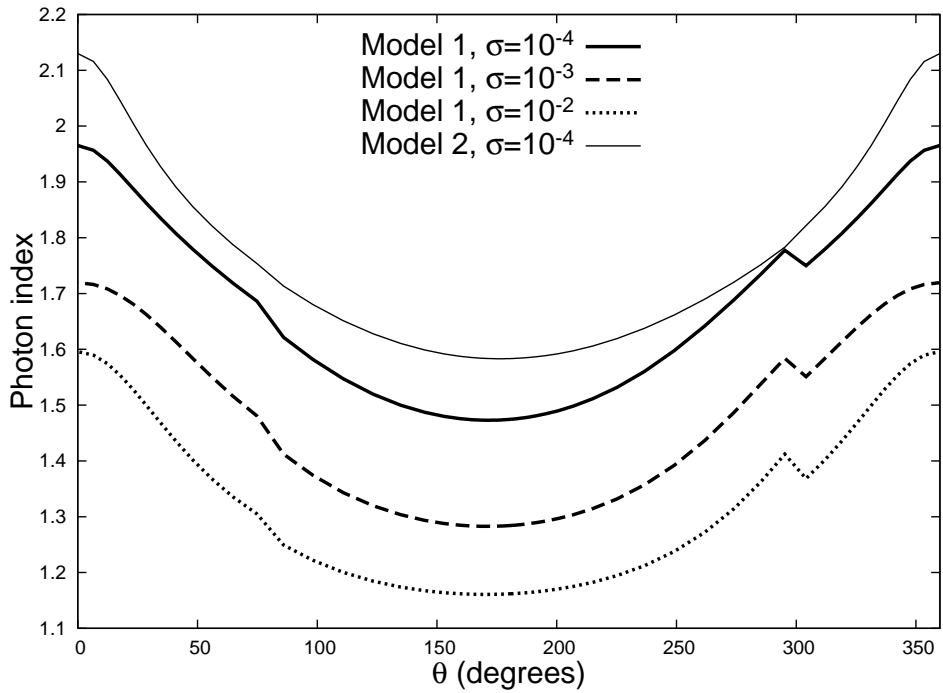}
\caption{Variation of the photon index in the 1-10~keV energy band with orbital phase. 
The lines denote the same model parameters as in Figure~\ref{synflux}.}
\label{photon}
\end{figure}

Figure~\ref{photon} shows the 
photon index of the calculated spectrum from the particles for $p_1=2.5$ 
in the 1-10 keV energy bands as 
a function of the orbital phase. The lines correspond to the same
 parameters as in  Figure~\ref{synflux}. 
We find that the photon index become smaller than 1.5 even though 
we have used  the  particle distribution with the index larger than $p_1>2$. 
The photon index $\alpha$, defined by $dN/dE\propto E^{-\alpha}$, 
 decreases as the pulsar moves from the apastron to the periastron, 
as can be seen from Figure~\ref{photon}. 
For example, the synchrotron emission in the 1-10 keV energy band has
a photon index $\alpha\sim 1.7$ at apastron for $\sigma=10^{-3}$  
(dashed line). This photon index follows from the relation $\alpha=(p_1+1)/2$ 
in the slow cooling regime with $p_1=2.5$,  which  
has been assumed in this section. On the other hand, a very hard spectrum 
with the photon index $\alpha\sim 1.3$  is obtained at the periastron, 
reflecting the fact that the photon index in the 1-10 keV energy band 
is related to the energy of the lower spectral break (see \S 1). The energy of 
the lower spectral break  is estimated from 
$E_1\sim 3he\Gamma_1^2B\sin\theta_p/4\pi m_ec$. For $\sigma=10^{-3}$ and 
the shock distance determined with $\dot{M}_p/f_{\Omega}=10^{-8}M_{\odot}$~/yr 
and  $\dot{M}_e/f_{\Omega,e}=10^{-7}M_{\odot}$~/yr (Model~1), the strength 
of the magnetic field is $B_2\sim 0.01$~Gauss at apastron 
and $B_2\sim 0.2$~Gauss at periastron. The 
energies, $E_1$, at apastron and periastron are 
$\sim 1.3$~keV and $\sim 26$~keV, where the energies are calculated with 
an average pitch angle $\theta_p=\sqrt{2/3}$. We find that 
near periastron, the energy of the lower spectral break is greater than 
10~keV.  As a result, the spectral behavior in the 1-10 keV energy band is 
described by the tail of the spectrum of the synchrotron radiation. 
In such a case,  we obtain a photon index 
smaller than $\alpha=1.5$, finding that 
it is possible to attain the photon index less than 1.5 with $p_1>2$.

We note that the spectrum in the 1-10 keV energy band becomes soft 
near the periastron in the model of Tavani \& Arons (1997). Based upon 
the adoption of the  approximate formula $\dot{\Gamma}_{IC}\propto \Gamma^2$ 
for cooling, the time scale of the inverse-Compton loss for particles with 
Lorentz factors $\Gamma\ge 10^6$ becomes smaller than that associated with 
adiabatic losses near periastron. Hence, the strong inverse Compton cooling 
softens the particle energy distribution and the synchrotron spectrum in the 
1-10~keV energy band such that $\alpha\sim 2$ near periastron. In contrast, 
we find that with the full functional form of the scattering cross section, 
adiabatic losses are more significant than the inverse-Compton losses for 
the entire orbit.  As a result, a softer spectrum near the periastron is not 
found in our computations, unless the pulsar wind parameters at the shock 
change with the orbital phase.

\subsection{Very high-energy emissions above 380~GeV}
\label{very}
\begin{figure}
\epsscale{.80}
\plotone{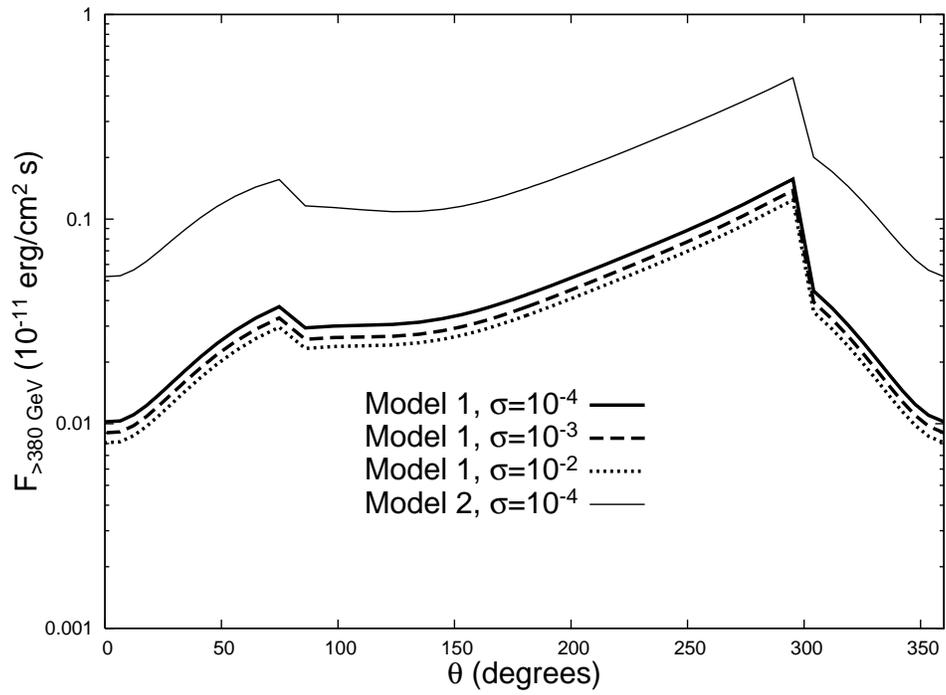}
\caption{Variation of the integrated flux above  380~GeV
 with respect to orbital phase.  
The lines correspond to same case as Figure~\ref{synflux}.}
\label{TeVflux}
\end{figure}

The calculated flux integrated above 380~GeV emitted by the inverse Compton 
process is shown as a function of the orbital phase in Figure~\ref{TeVflux}. 
The line types denote the same parameters as in Figure~\ref{synflux}. 
 The flux of the 
very high energy emission is not significantly affected by the value of
the $\sigma$ parameter, as we find from the  thick-solid, dashed and dotted 
lines.  However, upon comparison of the thick (Model~1) 
and solid (Model~2) lines we find that integrated fluxes are sensitive to 
  the shock distance for a given orbital phase. 
For example, the calculated fluxes at apastron differ by about a factor of 
5 even though the shock distances differ by only a factor of 2 between 
Model~1 and Model~2.
This can be understood from the 
approximate form for the luminosity of the inverse Compton process given by 
 $L_{IC}\propto \sigma_{IC} n_2 V N_{sp}$ with 
$N_{sp}$ corresponding to the number density of soft photons. If the 
luminosity $L_{IC}$ is recast as a function of the shock distance, we 
obtain $L_{IC}\propto r_s/(d-r_s)^2$ with $d$ corresponding to the orbital 
separation. At apastron ($d\sim 10$~AU), the shock distance
from  the pulsar (Figure~\ref{shockrad}) is $r_s=3.1$~AU for Model~1 
and $r_s=5.9$~AU for Model~2. The ratio of the fluxes becomes 
$L_{IC}(Model~2)/L_{IC}(Model~1)\sim 5.4$, which is consistent with 
the ratio of calculated fluxes for Model~1 (thick-solid line) 
and for Model~2 (thin-solid line) in Figure~\ref{TeVflux}.

It can be also seen that  the very high energy emission 
attains a maximum value at the epoch after the periastron, although 
the stellar photon energy density attains a maximum value at the periastron 
as the dashed-line in Figure~\ref{energy} shows.  
That is, the temporal variation with 
orbital phase is not symmetric with respect to the periastron. 
In the interaction between the pulsar  and equatorial stellar winds, 
the outflow velocity relative to the pulsar prior to periastron 
($80^{\circ}\le\theta\le 180^{\circ}$)
 is $v_{re}\sim\sqrt{v_{orb}^2+v_{w}^2}$, 
where $v_{w}$ is the velocity of the equatorial wind taken from
equation of (\ref{ewind}), and $v_{orb}$ is the orbital velocity of
pulsar. In this case the relative velocity $v_{re}$ is 
nearly constant before the periastron. 
 From equations (\ref{balance}) and (\ref{ewind}), we obtain 
the relation between 
the orbital separation and the shock distance, $d-r_s\propto r_s^{5/6}$
 with $m=0.4$.
 The flux of the inverse Compton radiation increases,  
$L_{IC}\propto r_s/(d-r_s)^2\propto r_s^{-2/3}$, as the pulsar approaches to 
the periastron. After the periastron ($180^{\circ}\le\theta\le300^{\circ}$),
 the direction of radial motion
 of the pulsar relative to the main star is positive so that 
  $v_{re}\sim\sqrt{|v_{orb}^2-v_{w}^2|}$. 
We can see that the orbital distance and the shock distance 
after the periastron 
is related as $d-r_s\propto r_s^{0.3}$. As a result, 
we obtain $L_{IC}\propto  r_s^{0.4}$.  Thus, the inverse-Compton luminosity
increases with  the shock distance after the periastron in contrast 
to the case prior to periastron. 
  This increase is due to the rapid change in shock distance for a 
small change of the orbital separation after the periastron, 
resulting in an increase
 in the emission volume with a slight decrease of the number density 
of the background soft photon field. This model temporal behavior provides  
a possible explanation of the observed fact that  
the flux of the very high energy photons attains a maximum 
value after the periastron, $\theta\sim 270^{\circ}$ (\S~\ref{comp}).

\begin{figure}
\epsscale{.80}
\plotone{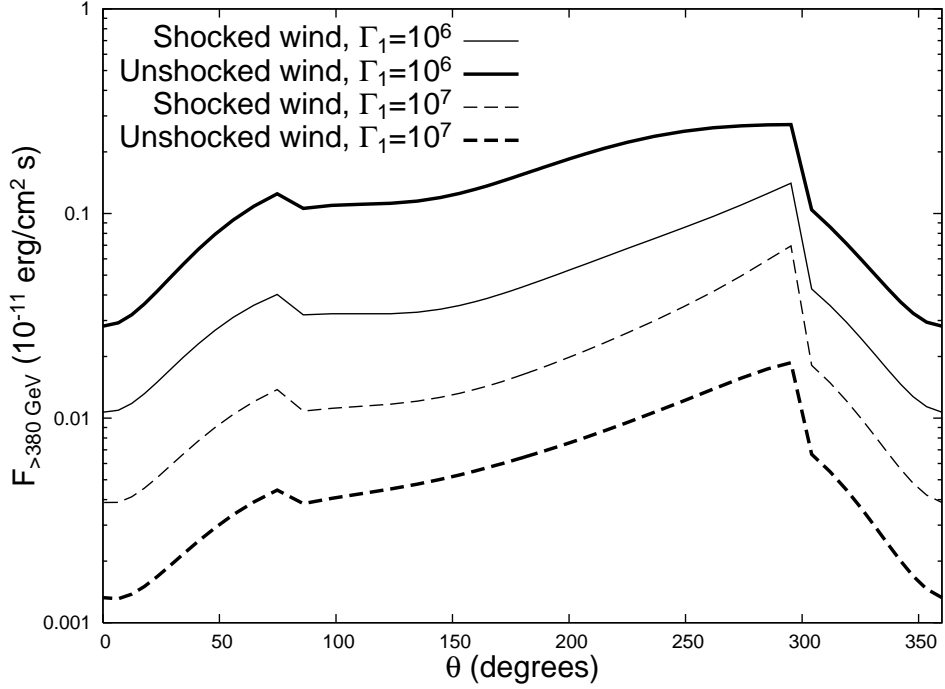}
\caption{Comparison of the integrated fluxes of the very high-energy radiation from 
the unshocked (thick lines) and shocked (thin lines) wind as a function 
of the orbital phase. The solid and dashed line are 
the results for the mono-energetic distribution of the 
unshocked wind with the Lorentz factor of $\Gamma=\Gamma_1$. The 
solid and dashed lines correspond to Lorentz factors
of $\Gamma_1=10^6$ and $10^7$, respectively.
All results are for $\sigma=10^2$, 
 $p_1=2.5$ for the shocked particles, and 
 $\dot{M}_p=10^{-9}M_{\odot}/f_{\Omega,p}$~/yr
and  $\dot{M}_e/f_{\Omega,p}=10^{-8}M_{\odot}$~/yr. }
\label{TeVflux2}
\end{figure}

The emission from the inverse Compton process of the unshocked wind can also 
contribute to the very high-energy emission. Figure~\ref{TeVflux2} compares 
the  temporal behaviors of integrated flux  above 380~GeV from the 
unshocked (thick lines) and shocked (thin lines) particles. In this 
comparison, we have used a mono-energetic distribution for the unshocked particles.  
  The solid and dashed lines are results for the 
Lorentz factor $\Gamma_1=10^6$ and $\Gamma_1=10^7$, respectively.
We find that the  integrated flux above 380~GeV from the
unshocked and shocked winds depends on $\Gamma_1$, decreasing with increasing Lorentz
factor because the inverse Compton scattering of the pulsar wind with the black body 
radiation ($kT\sim 2$~eV) from the Be star takes place in Klein-Nishina regime.
As $\Gamma_1 kT$ with $\Gamma_1=10^6$ or $10^7$ is greater than the rest mass energy 
of the electrons, the wind particles mainly upscatter the photons in the Rayleigh-Jeans 
region. Hence, the energy of incident stellar photon must lie further out in the 
Rayleigh-Jeans tail, corresponding to a lower photon density, to produce the very high 
energy photons in the case of higher Lorentz factors. Most of the soft photons above 
$m_ec^2/\Gamma_1$ are not scattered and, as a result, the number of scattered background 
photons, and therefore the integrated flux above 380~GeV, decreases 
with an increase of the Lorentz factor.

The 
integrated flux level from the unshocked wind is more sensitive to the Lorentz factor 
than from the shocked wind, as can be seen by comparing  the integrated fluxes above 380~GeV of the unshocked (thick-lines) and 
the shocked (thin-line) wind in Figure~\ref{TeVflux2}. 
This is a direct consequence of our assumption of a 
mono-energetic distribution of particles for the unshocked wind and a broken power law 
distribution for the shocked wind. In this case, the integrated flux from the 
unshocked wind can be larger than from the shocked wind. 

\begin{figure}
\epsscale{.80}
\plotone{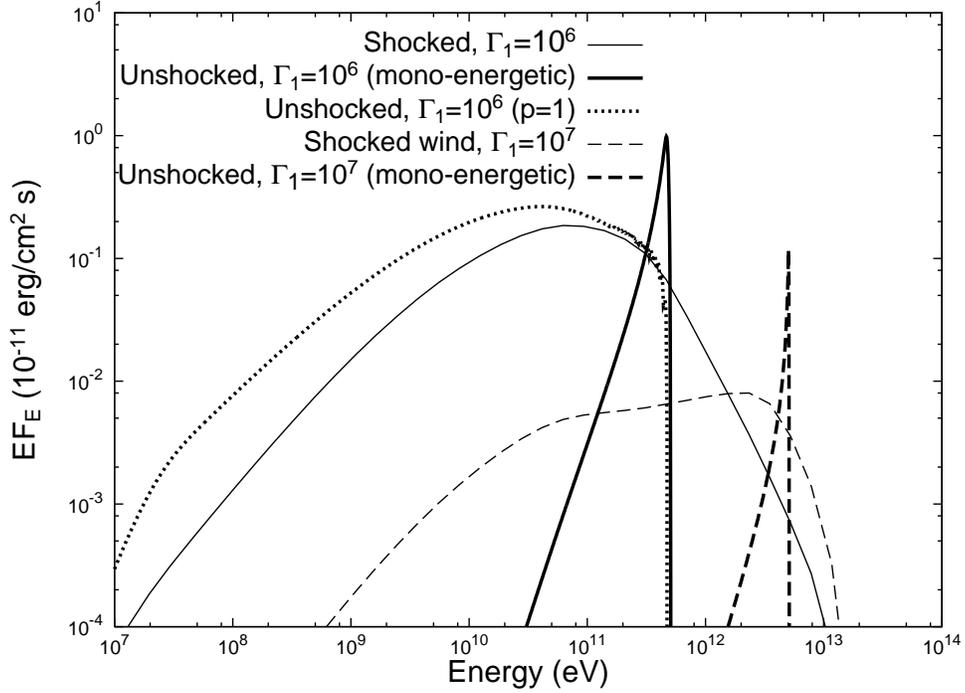}
\caption{Comparison of the spectrum of the very high-energy radiation from 
the unshocked (thick lines) and shocked wind (thin line) at the periastron.
The results are for a mono-energetic distribution of the 
unshocked wind with a Lorentz factor of $\Gamma=\Gamma_1$. The 
solid and dashed lines correspond to the Lorentz factor 
of $\Gamma_1=10^6$ and $10^7$, respectively. The dotted line represents 
the spectrum of the inverse-Compton process of the unshocked particles 
distribution with a power law index $p=1$ at $10^3<\Gamma<10^6$.
All results are for $\sigma=10^{-2}$, 
 $p_1=2.5$ for the shocked particles, and 
 $\dot{M}_p=10^{-9}M_{\odot}/f_{\Omega,p}$~/yr
and  $\dot{M}_e/f_{\Omega,p}=10^{-8}M_{\odot}$~/yr. 
}
\label{TeVspec}
\end{figure}

Similar temporal behavior of the integrated flux of 
these two components are seen in Figure~\ref{TeVflux2}, implying that 
these two components can not
 be distinguished based solely on the temporal behavior of the 
integrated flux. Furthermore, if 
the unshocked particles are distributed with a power law index, 
it will be difficult to distinguish the individual contributions, as 
dotted-line and thick-solid-line in Figure~\ref{TeVspec} indicate.  
Here, the dotted-line is the spectrum of the inverse-Compton process of 
the unshocked particles described by a power law index $p=1$ at $10^3\le 
\Gamma \le 10^6$.  In this case, the particle number is uniformly distributed 
in the energy space, but most of the kinetic energy of 
the flow is carried by the particles with the Lorentz factor of
 $\Gamma=10^6$ (see \S~\ref{dyunshock}). On the other hand,
it may be possible to distinguish
 the individual contributions if the spectral shape is measured and if 
the distribution of the unshocked particles is described as mono-energetic.
The spectrum associated with the inverse-Compton
 process of the unshocked wind with  the mono-energetic particle
 distributions are represented by the thick-solid line for  $\Gamma_1=10^6$  and by 
the thick-dashed-line for  $\Gamma_1=10^7$.
For an unshocked wind described by a mono-energetic distribution, 
the spectral shape 
of the inverse Compton emission has a line type structure, 
implying that the width of the spectrum of the high energy emission from the unshocked wind is much narrower than 
that from the shocked wind as already pointed out by 
Khangulyan et al. (2007) and Cerutti et al. (2008).  
 A higher spectral resolution study will be required  
to distinguish their relative importance. For example, the solid lines 
in Figure~\ref{TeVspec} imply that a spectral resolution $\delta E/E< 
40$~\% at 500~GeV for $\Gamma_1=10^{6}$ is required to distinguish between 
the emission from an unshocked and shocked pulsar wind.

\section{Discussion}
\label{discus}
\subsection{Comparison with the observations}
\label{comp}
The  X-ray observations (Hirayama et al. 1996; Chernyakova et al. 2006; 
Uchiyama et al. 2009)   
provide the temporal behavior of  the radiation spectrum and the photon 
index for the entire orbital phase. We refer to the orbital phases, where the 
X-ray data is available, as X1, X2, S1, A2, etc., following 
Chernyakova et al. (2006) and Uchiyama et al. (2009). 
The observed integrated flux in the 1-10~keV 
energy band increases by about an order of magnitude between 
the orbital phase $\theta\sim 50$~degree (X6) and $\theta\sim 100$~degree (S4).
If the pulsar wind properties ($\sigma$, $\Gamma_1$ and $p_1$) 
do not change with the orbital phase, the integrated X-ray flux in 
Figure~\ref{synflux} increases by only about factor two at the phase interval  
between  $\theta\sim 50$~degree and $\theta\sim 100$~degree, implying 
the pulsar wind properties are a function of the orbital phase. 

Uchiyama et al. (2009) fitted the X-ray  data of -15~days
 ($\theta\sim 95^{\circ}$), +30~days ($\sim 291^{\circ}$) and 
+618~days ($\sim 360^{\circ}$) with the synchrotron 
and inverse Compton radiation model, in which   
the adiabatic energy loss rate of the particles 
and the shock distance from the pulsar were parameterized. 
A similar computation with a parameterized adiabatic loss rate 
was carried out by Khangulyan et al. (2007) to explain 
the temporal behavior of the very high energy observations.

As an alternative interpretation of the temporal behavior, one can consider 
the magnetization parameter, the Lorentz factor $\Gamma_1$ 
and/or the power law index $p_1$ of the shocked particles to vary with orbital phase.
In fact, it is expected that $\sigma$ at the shock 
 is a function of the orbital phase because the shock distance from the pulsar varies. 
In addition, the terminal Lorentz factor, $\Gamma_1$, may also vary since it depends 
on the multiplicity factor, $\kappa$, (see equation (\ref{termL})), 
which can take different values for the different magnetic field lines 
accelerating the pulsar wind.  In the simple 
model investigated by  Daugherty and Harding (1996), for example, 
the multiplicity varies 
about factor of ten on the magnetic field lines coming from  
the polar cap region.
Therefore, the fitted Lorentz factor can also be a function of orbital phase, 
if the observed emission in different orbital phases emanates from
different field lines.

We compare the calculated flux and photon index in 1-10~keV bands 
with the observational results in the literature (eg. Uchiyama et al. 2009). 
We use the minimized  $\chi^2$ method to fit our calculated spectra 
in the 1-10~keV energy band with a single power law,  namely, 
 $\alpha=(N\sum x_iy_i-\sum x_i \sum y_i)/[N\sum x_i^2-(\sum x_i)^2]$, 
where  $N$ is the 
number of the sampling points  
in the energy range $1~\mathrm{keV}<E<10~\mathrm{keV}$, and $x_i$ 
and $y_i$ are the logarithm  of the energy and the number flux  of photons 
 of i-th sample, respectively. The shock distances from the pulsar were 
taken as an average of Model~1 and Model~2 shown in Figure~\ref{shockrad}.

\begin{figure}
\epsscale{.80}
\plotone{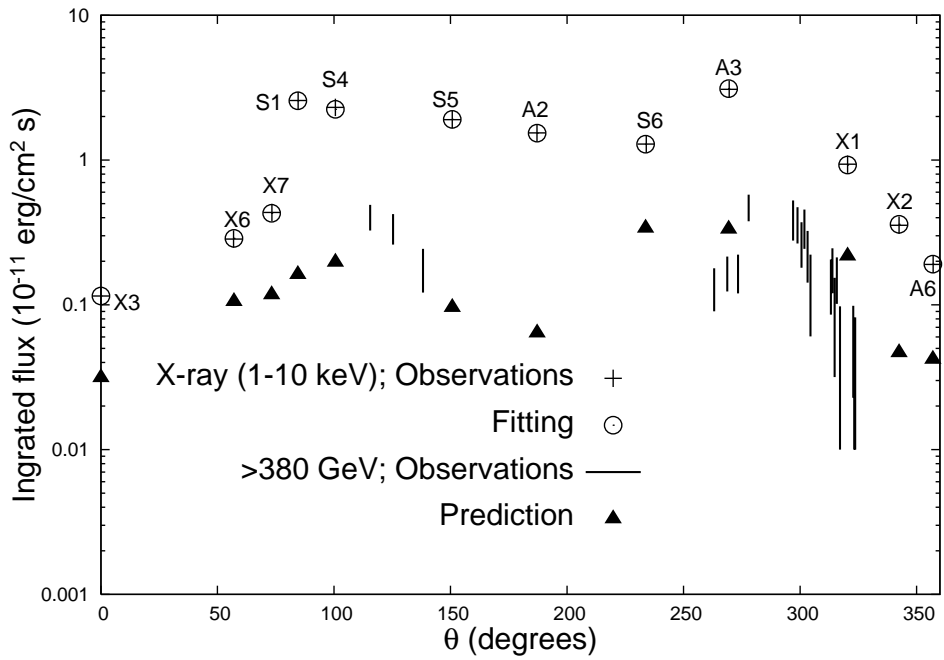} 
\caption{The temporal behavior of the integrated fluxes in the 1-10 keV energy band
and above 380~GeV. The cross symbols are results of the fitting for 
the integrated flux in the 1-10 keV energy band. The filled triangles denote
the predicted integrated flux above 380~GeV. The vertical lines represent 
the results of observations by H.E.S.S. (Aharonian et al. 2005).}
\label{Fit1}
\end{figure}

\begin{figure}
\epsscale{.80}
\plotone{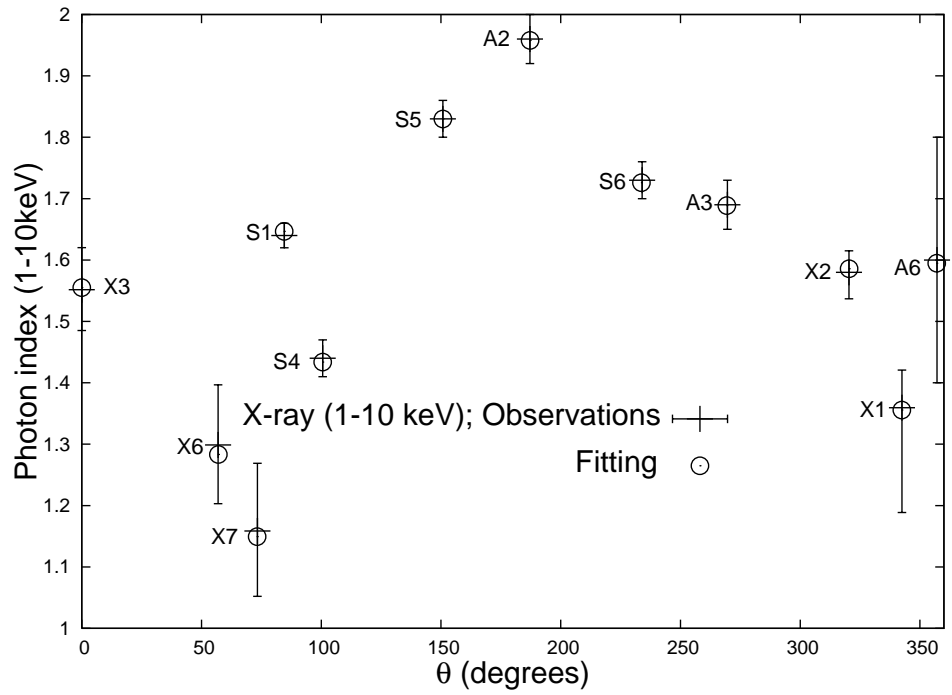}
\caption{The temporal behavior of the photon index of the spectrum 
in the 1-10~keV energy band.}
\label{Fitph1}
\end{figure}

An example of the fit parameters is summarized in 
Table~1, showing $\sigma$ and $\Gamma_1$ with $p_1=3$ 
at different orbital phases.  
 Figure~\ref{Fit1} shows the temporal behavior of the
 integrated flux in the 1-10~keV energy band and above 380~GeV. 
The temporal behavior of the photon index in the 1-10~keV energy band is shown 
in Figure~\ref{Fitph1}.
Considering the X-ray data near the apastron, we adopt $\sigma_0=8\times 
10^{-4}$ at orbital phase X3 as shown in Table~1.  Given this choice, a 
Lorentz factor $\Gamma_1=8.5\times 10^6$ provides a good fit to the observed 
photon index at this phase.  In this case, the calculated flux at X3 
is adjusted to the data with a normalization factor ($\sim 3$). 
 With this normalization factor, the integrated flux and photon 
index are fit to the X-ray data at other orbital phases, determining both 
$\sigma$ and $\Gamma_1$.

We note that the value of the $\sigma$ parameter at X3 is not arbitrarily 
chosen since a smaller value (e.g. $\sigma_0=10^{-4}$) 
leads to a calculated flux 
that is too small, requiring a normalization factor that is greater than a 
factor of ten. On the other hand, a larger value (e.g. 
$\sigma_0=5\times 10^{-3}$) 
does not produce a fit to the data near the periastron (A3) for a wide 
range of $\sigma$ and $\Gamma_1$ combinations.  For a shock distance 
corresponding to an average value determined in Models~1 and 2, we find that 
$\sigma_0\sim 8\times 10^{-4}-10^{-3}$ 
is suggested by the data in the 1-10~keV energy band. 

The results in Table~1 are based on an assumed power law index ($p_1=3$) for 
the particle energy distribution (see equation (\ref{power})) for the entire 
orbit. Such a choice was based on the fact that (i) the observed
photon indices 
are greater than 1.7 at and near the apastron and periastron, excluding a 
particle energy distribution with a very hard index, and (ii) such 
a choice is required to explain the observed soft spectrum $\alpha \sim 1.95$ 
in the 1-10~keV energy band at orbital phase A2.  We note that the observed 
very hard spectrum $\alpha\sim 1.2$ at phase X7 is  obtained with the 
power law index $p_1=3$ because of (i)
 a spectral break larger than $E_c>10$~keV with 
the fitted Lorentz factor of  $7.8\times 
10^6$ and (ii) the very hard spectrum with $p_2\sim 0.45$ below $E_c$. 

Based on the inferred parameters of the pulsar wind deduced by the X-ray 
observations and without any additional free parameters, the predicted very high energy 
emission above 380~GeV (see filled triangles in Figure~\ref{Fit1}) is found to be 
qualitatively consistent with the observations (vertical solid lines), except 
at orbital phase S6 where the predicted flux may be too large by comparing to a value 
extrapolated from the present H.E.S.S. observations.  

The predicted flux above 380~GeV attains a maximum value after 
the periastron as discussed in \S~\ref{very}. 
LS~I+$61^{\circ}303$ also shows a similar temporal behavior 
to the PSR B1259-63/SS2883 system in the very high energy bands  
as the flux attains a maximum value after periastron.  However, the cause 
is different from that in PSR B1259-63/SS2883 system. In particular,  
Sierpowska-Bartosik \& Torres (2009) argue that the observed 
temporal behavior of  LS~I+$61^o303$ is  caused by 
the  absorption of the photons near periastron, where 
the optical depths greater than unity are expected for the very high energy 
photons propagating to the observer. Therefore most of high-energy photons 
emitted near the periastron are absorbed by the stellar photons for 
LS~I+$61^{\circ}303$.  In contrast, the optical depth of the photons for 
the PSR B1259-63/SS2883 system  is smaller than unity 
for entire orbit (see Figure~\ref{depth}).

The solid line in Figure~\ref{average} represents the model spectrum for the 
shocked particles averaged during the 4-month time interval from -20~days 
($\theta\sim 83^{\circ}$) and +100~days ($\theta\sim 324^{\circ}$), which 
was chosen for comparison with the four month observations (February - May in 
2004) by H.E.S.S. We find that the predicted level of the flux is consistent 
with the H.E.S.S. observations below 1~TeV (Aharonian et al. 2005), while  
the model predicts a softer spectrum than observed above 1~TeV. On the other 
hand, we find that detection by the Fermi telescope is not expected at the 
GeV energy bands for this model. Note that the calculated  spectrum averaged 
during the four month time interval -60~days ($\theta\sim 48^{\circ}$) and 
+60~days ($\theta\sim 312^{\circ}$) is very similar to and can not be 
distinguished from the solid line in Figure~\ref{average}.

In summary, our phenomenological analysis places constraints on the parameters 
$\sigma$ and $\Gamma_1$ as a function of orbital phase in the $\gamma$-ray binary 
system PSR B1259-63/SS 2883.  We have found that $\sigma$ tends to increase as the 
pulsar companion to the Be star approaches periastron (orbital phase A2), suggesting 
that the energy conversion process from the magnetic field to bulk motion of the 
pulsar wind varies on the scale of the shock distance in the binary.
In addition, constraints on $\Gamma_1$ have been inferred from the hardness of 
the spectrum.  The hardening seen in the observed spectrum in Figure~\ref{Fitph1} 
between orbital phases X3 and X7 can be fit for Lorentz factors $\Gamma_1$ which 
do not significantly vary. 
However, after orbital phase X7, the observed photon index returns to a value 
steeper than the photon index 1.2 at X7 (Figure~\ref{Fitph1}), which is not 
easily produced by the synchrotron model for a fixed $\Gamma_1$ throughout 
the orbit. As a result, the Lorentz factors near the periastron 
($\Gamma_1\sim 3\times 10^5$ at A2) are more than an order of magnitude 
smaller than that near apastron  ($\Gamma_1\sim 8\times 10^6$ at X3). 
The large variation of the fitted Lorentz factor $\Gamma_1$ 
may imply that the observed emission in various orbital phases emanate 
from the different field lines giving rise to a range of multiplicities.  The 
Lorentz factors from 
$\Gamma_1=8\times 10^6$ to  $3\times 10^6$ correspond to 
the multiplicities of $\kappa\sim 3\times 10^3$ and $10^2$, respectively 
(see equation (\ref{termL})) and are in the range predicted by 
pulsar models (Daugherty \& Harding 1996; Hibschman \& Arons 2001). 
Some support for this picture is provided in the simple model by Daugherty \& Harding 
(1996), which indicates that the multiplicity changes by about a factor of 10 
across the magnetic field lines emanating from the polar cap region (see 
figure~3 in their paper). The predicted variation for $\Gamma_1$ ($\sim30$) 
in the present paper is larger than indicated in this model, but may be in 
an acceptable range.

Although the above fits are in reasonable accord with the observations, our lack of 
a physical understanding for the inferred parameter values suggests that an alternative interpretation for the observed temporal behavior be considered in order to determine 
the sensitivity of the fits to the particular parameterizations. Specifically, we examine 
the possibility that the power law index $p_1$ of the accelerated particles varies 
throughout the orbit.  The present shock acceleration models (e.g. Baring 2004)  predict a
value in the range $1.5\la p_1\la 3$, which can accommodate 
the observed  ranges of the photon indices $1.2\la\alpha\la2$. 
Therefore, the X-ray data were fitted with a model parameterizing 
$\sigma$ and $p_1$. The Lorentz factor was fixed at $\Gamma_1=5\times 10^5$, which 
was chosen to follow  $\Gamma_1=4.5\times 10^5$ obtained in Uchiyama et al. (2009),  for 
the entire orbital phase.  Table~2 lists the fitted $\sigma$ parameters and power law 
indices  $p_1$ with the orbital phases. 

It can be seen that the fitted power law indices, $p_1$, listed in Table~2 
are less than $p_1=3$ chosen for the fits in Table~1 in most phases of the orbit.
For Table~1, $p_1=3$ was chosen to obtain the observed softest photon index 
$\alpha\sim2$ near periastron (phase A2 in Figure~\ref{Fitph1}) and harder 
spectra were produced by adjusting the Lorentz factors $\Gamma_1$. Here, the 
synchrotron radiation spectral break, corresponding to the break of the 
particle distribution at $\Gamma_1$, was located at energies larger than 
1~keV for most of the orbital phase. In Table 2, on the other hand, the fits 
using an adopted Lorentz factor of $\Gamma_1=5\times 10^5$ produce a spectral 
break below 1~keV for the entire orbital phase. This indicates that the 
computed photon index in the 1-10~keV energy band behaves 
as $\alpha= (p_1+1)/2$, for X-ray emission via synchrotron radiation process 
in the slow cooling regime. 
As a result, power law indices less than $p_1=3$ can produce photon index fits 
less than $\alpha<2$ observed most of the orbital phase.

The solid line in Figure~\ref{average1} represent
 the model spectrum  averaged during 
the orbital phases from -20~days ($\theta\sim 83^{\circ}$) and +100~days 
($\theta\sim 324^{\circ}$).
Similar to Figure~\ref{average}, the model spectrum in the TeV band is 
found to be qualitatively consistent  with the observations below 1~TeV, while 
the model spectrum is softer than observations above 1~TeV. On the other hand, 
the emission from the shocked particles in the GeV band
is expected to be detected by the $Fermi$ telescope in such a model. 
For comparison with the solid line, we plot the model spectrum averaged during 
the four month time interval but from -60~days ($\theta\sim 
48^{\circ}$) to  +60~days ($\theta\sim 312^{\circ}$). Although the spectral 
shapes for the two time intervals are similar to each other, 
the level of the fluxes can be observed in Figure~\ref{average1}, in contrast to 
the case exhibited in Figure~\ref{average}.

\subsection{Radio emission}

We have not attempted to compare the radio emission from our model with the observations 
since the model for the non-pulsed radio emission is more complex than that described for 
the $X/\gamma$-ray emission. In particular, it is expected that the radio waves are emitted 
by particles cooled to low energy $(\Gamma\sim 10^{2}-10^{3}$) over a volume which is large
compared to the emitting region of photons characterized by higher energies.  
That is, the properties of the radio emission 
are more sensitive to the cooling process, the shock geometry, and the propagation of the 
shocked wind than the emission in the X/$\gamma$-ray regime. Hence, studying the radio 
emission as a means to probe the properties of the shocked and unshocked pulsar wind 
is not as direct as for the high energy emission because the radio emitting particles no 
longer characterize the properties of the pulsar wind due to the radiation cooling 
process. A model describing the radio emission from the $\gamma$-ray
 binaries has been presented in Dubus (2006b).

\subsection{Future perspective}
\label{perspect}
\begin{figure}
\epsscale{.80}
\plotone{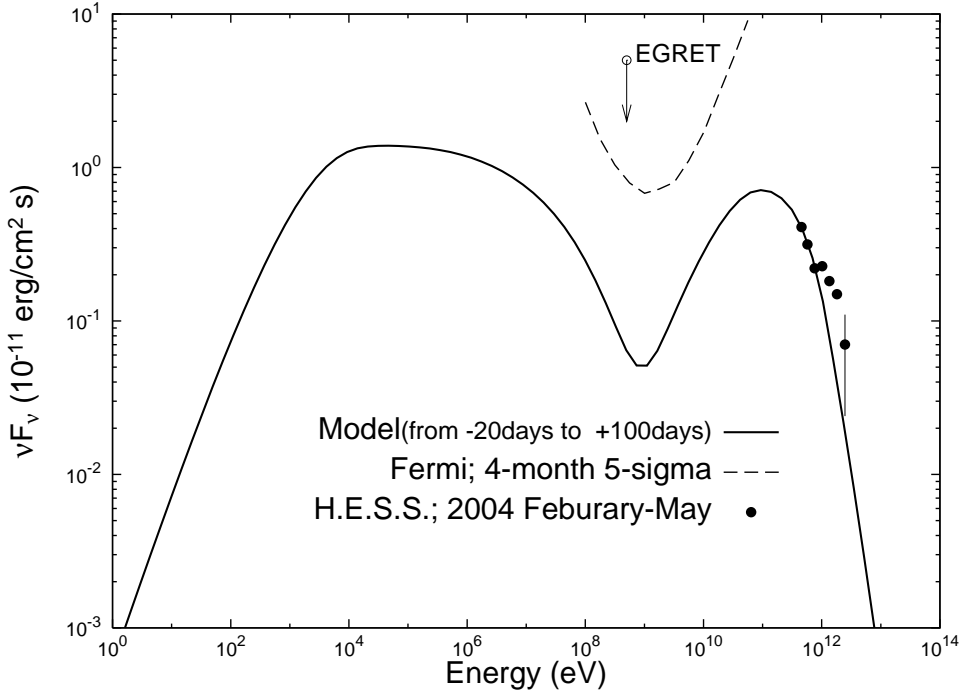}
\caption{The averaged spectrum of the emission from the shocked wind 
during the 4 month time interval from -20~days ($\theta\sim 83^{\circ}$)
 and +100~days ($\theta\sim 324^{\circ}$). The spectrum is calculated  
with the model in which $p_1=3$ is fixed, and 
$\sigma$ and $\Gamma_1$ vary with the orbital phase. The dashed-line 
represents the sensitivity for a 4-month observation using the $Fermi$,  
 $\mathrm
{http://www-glast.slac.stanford.edu/software/IS/glast\_lat\_performance.htm}$. 
The filled circle is results of the observation done by H.E.S.S. 
during February-May, 2004 (Aharonian et al. 2005). 
Note that the  calculated  spectrum averaged during the four 
month time interval -60~days ($\theta\sim 
48^{\circ}$)  and +60~days ($\theta\sim 312^{\circ}$) is 
 very similar to and can not be distinguished from the solid line. 
An upper limit at $\sim$500~MeV is obtained from EGRET observations.} 
\label{average}
\end{figure}

\begin{figure}
\epsscale{.80}
\plotone{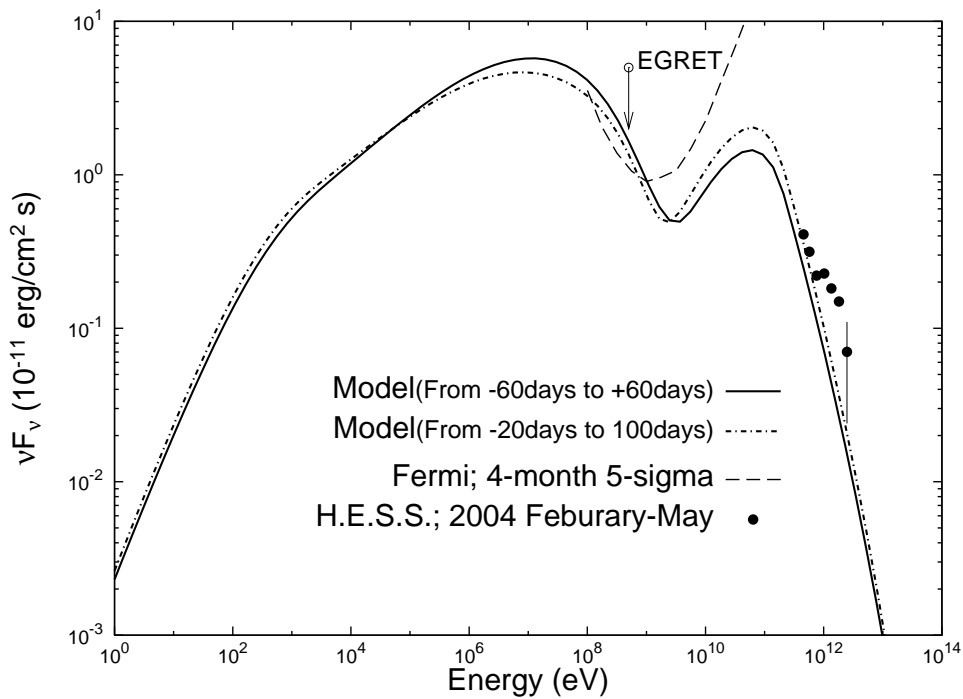}
\caption{The averaged spectrum of  the emission from the shocked wind 
during a 4 month time interval from -20~days ($\theta\sim 83^{\circ}$)
 and +100~days ($\theta\sim 324^{\circ}$) and -60~days ($\theta\sim 
48^{\circ}$) to  +60~days ($\theta\sim 312^{\circ}$). 
 The spectrum is calculated with the model in which $\Gamma_1=5\times 10^5$ is fixed, and 
$\sigma$ and $p_1$ vary with the orbital phase. 
An upper limit at $\sim$500~MeV is obtained from EGRET observations.} 
 \label{average1}
\end{figure}

\begin{figure}
\epsscale{.80}
\plotone{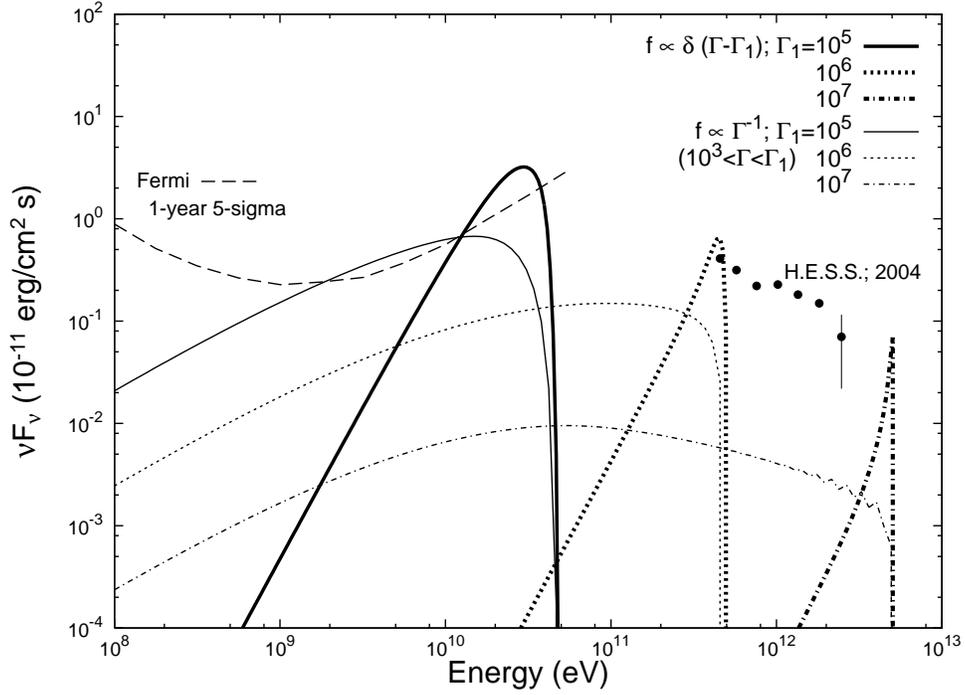}
\caption{The inverse-Compton spectra from the unshocked wind. 
The results are for the averaged spectra during 1 year of 
the periastron passage.
The thick and the thin lines represent the spectra from the particles with 
a mono-energetic 
distribution and a power law distribution with index $p=1$, respectively. 
The solid, dotted and dashed-dotted lines are results for $\Gamma_1=10^5$, 
$10^6$ and $10^7$.}
 \label{aveunshock}
\end{figure}

Future observations of PSR B1259-63/SS 2883 with the Fermi telescope will provide further 
constraints on the properties of the emitting particles, especially for 
the non-thermal emission in the MeV-GeV energy range.  For example, the Fermi 
telescope can constrain the electron energy distribution of the shocked wind. 
In comparing Figures~\ref{average} and ~\ref{average1}, the slope of the 
flux distribution in the keV-GeV energy bands and the level of flux in the MeV-GeV 
energy bands are sensitive to the adopted power law index $p_1$. In addition, 
the study of the temporal behavior by the $Fermi$ telescope can be compared 
with the predicted behavior of the flux in the 0.02-100~ GeV energy bands as 
summarized in Table~1 and Table~2.
Moreover, a measurement of a spectral break for emission associated with the 
synchrotron radiation, predicted in the MeV-GeV energy bands,
provides additional constraints 
on the highest energy of the accelerated particles at the shock. 
Fermi observations can also be used to investigate the unshocked 
wind as discussed in  Khangulyan et al. (2007). 
Figure~\ref{aveunshock} 
summarizes the spectra of the inverse-Compton process of the unshocked 
 particles  with a mono-energetic distribution (thick lines) 
and with a power law distribution ($p=1$) for different Lorentz factors $\Gamma_1$.
The various line types represent the results for the different Lorentz factor 
$\Gamma_1$ with the solid, dotted  and dotted-dashed lines corresponding 
to the results for $\Gamma_1=10^5$, $10^6$ and $10^7$. Here,
the calculated spectra are results for the averaged spectra during 
the three-month time interval at the periastron passage. 
We find from Figure~\ref{aveunshock}
 that the emission from the unshocked wind could be detectable by 
the Fermi telescope if most of the kinetic energy of the flow is carried by 
the particles with  the Lorentz factor of
 $\Gamma\sim 10^5$. 

We have shown that both the shocked and unshocked pulsar wind can contribute 
to the very high energy emission and that higher spectral resolution studies 
will be required to distinguish their relative importance. If the energy 
distribution of the unshocked wind particles are described by a power law, 
it is difficult to distinguish those two components. However, 
it may be possible to distinguish the individual contributions if 
the distribution of the unshocked particles is described as mono-energetic, 
which produces a line type structure of the spectrum for the inverse Compton emission.
 It is found that 
the width of the spectrum of the high-energy emission from the mono-energetic 
wind particles  becomes 
increasingly narrow for larger Lorentz factors, implying higher spectral resolution 
is required for larger $\Gamma_1$. For example, a spectral resolution 
$\delta E/E< 
40$~\% at 500~GeV for $\Gamma_1=10^{6}$ would be required to distinguish between 
the emission from the unshocked and shocked pulsar wind. 

Finally, we point out that the phenomenological approach adopted in this paper
has the potential to place constraints not only on the pulsar wind properties,
but also on the properties of the stellar wind from the Be star in SS 2883. 
For example, the index $m$ characterizing the velocity distribution of the equatorial 
outflow (see equation \ref{ewind}) in SS 2883 is not well constrained by the observations. 
However, the very high energy emission above 380~GeV is a function of $m$ with the flux 
decreasing for larger values of $m$ as a consequence of the closer proximity of the 
termination shock to the pulsar.  This follows from the fact that (i) the photon 
number density of the stellar radiation field decreases at the shock due to the smaller 
solid angle of the Be star as seen from the shock, and (ii) the high energy particles 
lose their energy faster via the synchrotron radiation with a stronger magnetic 
field at the shock. With these two effects, the very high energy emission above 380~GeV 
decreases for larger values of $m$. Specifically, taking $m=1.25$, which is used in 
Sierpowska-Bartosik \& Bednarek (2008) in contrast to $m=0.4$ adopted in this paper, 
the predicted  integrated flux above 380~GeV after periastron passage (at the orbital 
phase A3) is decreased by about a factor of 3. Hence, the dependence of the high energy 
integrated flux on the outflow index $m$ may provide an additional diagnostic to probe 
the outflow index $m$.  However, the uncertainties in the current observations preclude 
the usefulness of this diagnostic to discriminate the outflow indices at present. 
Therefore, observations at higher sensitivity as well as more sophisticated theoretical 
models will be necessary to realize this potential.

\acknowledgments
We wish to express our thanks to the referee for detailed comments which significantly improved 
the paper. In addition we also thank R.Huang for providing XMM-Newton and ASCA data. 
This work was supported by  the Theoretical Institute for Advanced Research in 
Astrophysics (TIARA) operated under Academia Sinica and the National Science Council 
Excellence Projects program in Taiwan, administered through
grant NSC 96-2752-M-007-007-PAE. 

%\section{Appendix material}

%% Use LXaTeX's thebibliography environment to mark up your reference list.



\clearpage

%% the file, uncomment the \plotone, \plottwo, and
%% If you need a layout that cannot be achieved with \plotone or
%% \plottwo, you can invoke the graphicx package directly with the
%% \includegraphics command or use \plotfiddle. For more information,
%% please see the tutorial on "Using Electronic Art with AASTeX" in the
%% documentation section at the AASTeX Web site,
%% http://www.journals.uchicago.edu/AAS/AASTeX.
%%
%% The examples below also include sample markup for submission of
%% supplemental electronic materials. As always, be sure to check
%% the instructions to authors for the journal you are submitting to
%% for specific submissions guidelines as they vary from

%% This example uses \plotone to include an EPS file scaled to
%% 80% of its natural size with \epsscale. Its caption
%% has been written to indicate that additional figure parts will be
%% available in the electronic journal.
\newpage

\newpage

\newpage


\newpage 

\begin{landscape}
\begin{table}[h]
\begin{tabular}{cc|cccccccccccc}
\hline\hline
\multicolumn{2}{c|}{Orbital phase } & \multicolumn{1}{c}{X3} & \multicolumn{1}{c}{X6} & X7 & S1 & S4 & A5 & A2 & S6& A3  & X1 & X2 & A6 \\
\cline{3-14}
Fitting parameters & $\sigma~(\times 10^{-3})$ & 0.8 & 1.2 & 2.2 & 2.1 & 11 & 11 & 15 & 6.2 & 70 & 9.2 & 3.3 & 1.7 \\
($p_1=3$) & $\Gamma_1 (\times 10^6)$ &8.5  & 7.6 & 7.8 & 1.1 & 1.7 & 0.57 & 0.3 & 1 & 0.8 & 2.8 & 8 & 6.5 \\
\hline
$F_{\mathrm{0.02-10^2GeV}}$& ($10^{-11}\mathrm{erg/cm^2 s})$ & 0.017 & 0.1 & 0.19 & 2.3 & 2 & 5.8 & 12 & 4.3 & 4.8 & 0.46 & 0.028 & 0.028 \\
\hline
\end{tabular}
\caption{The fitted $\sigma$ parameter and the Lorentz factor $\Gamma_1$. 
The results are for a power law index $p_1=3$ of the electron 
energy distribution and for the shock distance averaged of 
 Model~1 and Model~2 in Figure~\ref{shockrad}. The flux in the 0.02-100 GeV 
energy range is also listed.}
\end{table}
\end{landscape}

\begin{landscape}
\begin{table}[h]
\begin{tabular}{cc|cccccccccccc}
\hline\hline
\multicolumn{2}{c|}{Orbital phase } & \multicolumn{1}{c}{X3} & \multicolumn{1}{c}{X6} & X7 & S1 & S4 & A5 & A2 & S6& A3  & X1 & X2 & A6 \\
\cline{3-14}
Fitting parameters & $\sigma~(\times 10^{-3})$ & 2 & 5 & 40 & 40 & 80 & 9 & 7.5 & 7.9 & 10 & 25& 12 & 5 \\
($\Gamma_1=5\times 10^5$) & Index $p_1$ &2.1  & 1.62 & 1.45 & 2.35 & 1.99 & 2.8 & 3.2& 2.57 & 2.55 & 2.2 & 1.72& 2.2 \\
\hline
$F_{\mathrm{0.02-10^2GeV}}$& ($10^{-11}\mathrm{erg/cm^2 s})$ & 0.79 & 6.8 & 123 & 17 & 66 & 8.6 & 9.6 & 11 & 13 & 11 & 5.6 & 0.88\\
\hline
\end{tabular}
\caption{The fitted $\sigma$ parameter and the power law index  $p_1$. 
The results are for a Lorentz factor $\Gamma_1=5\times 10^5$ of the electron 
energy distribution and for the shock distance averaged of 
 Model~1 and Model~2 in Figure~\ref{shockrad}. The predicted flux in the 0.02-100 GeV 
energy range is also given.}
\end{table}
\end{landscape}

\newpage

\newpage

\newpage

\end{document}